\documentclass[preprint]{aastex}

\newcommand{\lya}{Ly$\alpha$}
\newcommand{\nv}{N\,{\sc v}}
\newcommand{\siii}{Si\,{\sc ii}}
\newcommand{\oi}{O\,{\sc i}}
\newcommand{\cii}{C\,{\sc ii}}
\newcommand{\siiv}{Si\,{\sc iv}}
\newcommand{\civ}{C\,{\sc iv}}
\newcommand{\heii}{He\,{\sc ii}}
\newcommand{\oiii}{O\,{\sc iii]}}
\newcommand{\aliii}{Al\,{\sc iii}}
\newcommand{\siiii}{Si\,{\sc iii]}}
\newcommand{\ciii}{C\,{\sc iii]}}
\newcommand{\mgii}{Mg\,{\sc ii}}
\newcommand{\feii}{Fe\,{\sc ii}}
\newcommand{\hb}{H$\beta$}

\newcommand{\civd}{C\,{\sc iv} $\lambda\lambda$1548,1550}
\newcommand{\feiiaa}{Fe\,{\sc ii} $\lambda$1608}
\newcommand{\aliia}{Al\,{\sc ii} $\lambda$1670}
\newcommand{\aliiid}{Al\,{\sc iii} $\lambda\lambda$1854,1862}
\newcommand{\feiiab}{Fe\,{\sc ii} $\lambda$2600}
\newcommand{\mgiid}{Mg\,{\sc ii} $\lambda\lambda$2796,2803}
\newcommand{\mgia}{Mg\,{\sc i} $\lambda$2852}

\newcommand{\mgit}{Mg\,{\sc i}}
\newcommand{\aliit}{Al\,{\sc ii}}

\begin{document}

\title{GEMINI NEAR-INFRARED SPECTROSCOPY OF LUMINOUS {\boldmath $z\sim6$} 
  QUASARS: CHEMICAL ABUNDANCES, BLACK HOLE MASSES, AND \mgii\ 
  ABSORPTION\footnote{
  Based on observations obtained at the Gemini Observatory (acquired through 
  the Gemini Science Archive), which is operated by the Association of 
  Universities for Research in Astronomy, Inc., under a cooperative agreement 
  with the NSF on behalf of the Gemini partnership: the National Science 
  Foundation (United States), the Particle Physics and Astronomy Research 
  Council (United Kingdom), the National Research Council (Canada), CONICYT 
  (Chile), the Australian Research Council (Australia), CNPq (Brazil) and 
  CONICET (Argentina).}}

\author{Linhua Jiang\altaffilmark{1}, Xiaohui Fan\altaffilmark{1},
  Marianne Vestergaard\altaffilmark{1}, Jaron D. Kurk\altaffilmark{2},
  Fabian Walter\altaffilmark{2}, Brandon C. Kelly\altaffilmark{1},
  and Michael A. Strauss\altaffilmark{3}}
\altaffiltext{1}{Steward Observatory, University of Arizona,
  933 North Cherry Avenue, Tucson, AZ 85721}
\altaffiltext{2}{Max-Planck-Institut f\"{u}r Astronomie, K\"{o}nigstuhl 17,
  D-69117 Heidelberg, Germany}
\altaffiltext{3}{Department of Astrophysical Sciences, Princeton University, 
  Princeton, NJ 08544}
\begin{abstract}

We present Gemini near-infrared spectroscopic observations of six luminous
quasars at $z=5.8\sim6.3$. Five of them were observed using 
Gemini-South/GNIRS, which provides a simultaneous wavelength coverage of 
0.9--2.5 $\mu$m in cross dispersion mode. The other source was observed in $K$
band with Gemini-North/NIRI. 
We calculate line strengths for all detected emission lines and use their 
ratios to estimate gas metallicity in the broad-line regions of the quasars. 
The metallicity is found to be supersolar with a typical value of $\sim4$ 
Z$_{\sun}$, and a comparison with low-redshift observations shows no strong 
evolution in metallicity up to $z\sim6$. The \feii/\mgii\ 
ratio of the quasars is $4.9\pm1.4$, consistent with low-redshift 
measurements. We estimate central BH masses of $10^9$ to $10^{10}$ M$_{\sun}$ 
and Eddington luminosity ratios of order unity. 
We identify two \mgiid\ absorbers with rest equivalent width 
$W_0^{\lambda2796}>1$ \AA\ at $2.2<z<3$ and three \mgii\ absorbers with 
$W_0^{\lambda2796}>1.5$ \AA\ at $z>3$ in the spectra, with the two most 
distant absorbers at $z=4.8668$ and 4.8823, respectively. The redshift number 
densities ($dN/dz$) of \mgii\ absorbers with $W_0^{\lambda2796}>1.5$ \AA\ are 
consistent with no cosmic evolution up to $z>4$.

\end{abstract}


\section{INTRODUCTION}

In recent years we have witnessed the discoveries of quasars at $z\sim6$; 
at this epoch the universe was less than one Gyr old 
\citep[e.g.][]{fan03,fan06,coo06,mcg06}. These high-redshift quasars are among 
the most luminous objects known and provide direct probes of the early 
universe when the first generations of galaxies and quasars formed. Quasar 
activity and the formation processes of galaxies and supermassive black holes 
(BHs) are closely correlated \citep[e.g.][]{kau00,wyi03,hop05,cro06}, so these 
$z\sim6$ objects are essential for studying the accretion history of BHs, 
galaxy formation, and chemical evolution at very early epochs. High-redshift 
luminous quasars harbor BHs with masses of $10^9$--$10^{10}$ M$_{\sun}$ and 
emit near the Eddington limit \citep[e.g.][]{bar03,ves04,jia06,kur07}, 
revealing the rapid growth of central BHs. Their emission line ratios show 
solar or supersolar metallicity, as is found in low-redshift quasars 
\citep[e.g.][]{bar03,die03a,fre03,mai03}, indicating that there was vigorous 
star formation and element enrichment in the first Gigayear.

Understanding quasars requires observations from X-ray to radio wavelengths;
each spectral region is due to different physical mechanisms and probes 
different regions of the active nucleus. In particular, the rest-frame UV 
spectrum, from the \lya\ emission line to the \feii\ bump at 2000--3000 \AA, 
contains strong diagnostic emission lines and provides key information on the 
physical conditions and emission mechanisms of the broad-line region (BLR). 
The bulk motions of the BLR can be used to determine the mass of the central 
BH, while chemical abundances in the BLR are important in understanding
the history of star formation in the host galaxy. Photoionization models show 
that various emission-line ratios provide reliable estimates of metallicity in 
the BLRs of quasars \citep[e.g.][]{ham02,nag06}. Strong permitted line ratios, 
such as \nv\ $\lambda$1240/\civ\ $\lambda$1549 (hereafter \nv/\civ),
\nv\ $\lambda$1240/\heii\ $\lambda$1640 (hereafter \nv/\heii), and 
\feii/\mgii\ $\lambda$2800 (hereafter \feii/\mgii), are relatively easy to 
measure even in $z\sim6$ quasars \citep[e.g.][]{ham93,ham99}.
Fe abundance is of particular interest \citep[e.g.][]{ham93,ham99}: most of 
the Fe in the solar neighborhood is generated by Type Ia supernovae 
(SNe Ia), whose precursors are believed to be long-lived intermediate-mass 
stars in close binaries, so appreciable Fe enrichment can only happen on a 
timescale of one Gyr after the initial starburst
\citep[e.g.][]{gre83}. On the contrary, the production of $\alpha$ elements
such as Mg and O is dominated by SNe of types II, Ib and Ic, which 
explode very soon after the initial burst. Therefore, the Fe/$\alpha$ ratio is 
expected to be a strong function of age in young systems, and thus it puts a 
useful constraint on the age and chemical enrichment of the gas in the quasar
environment.

At low redshift, elemental abundances measured from strong emission and 
intrinsic absorption lines have revealed that quasar environments have solar 
or supersolar metallicity. At $z\sim6$, measurement of these spectral lines 
is difficult, as the rest-frame UV waveband is 
redshifted to the NIR range. NIR spectroscopy has been obtained for a few 
$z\sim6$ quasars \citep[e.g.][]{goo01,pen02,bar03,fre03,mai03,ste03,iwa04}. 
These observations only provide NIR spectra with low signal-to-noise ratios 
(SNRs) or partial wavelength coverage. Even basic properties such as the 
average continuum slope are not well measured. Recently, \citet{kur07} 
obtained NIR spectra with high SNRs for a sample of five $z\sim6$ quasars with 
ISAAC on VLT. 
They used the ISAAC data to study the \feii/\mgii\ ratios, BH masses, and the 
Str\"omgren spheres around quasars. In this paper we present Gemini NIR 
spectroscopy of six $z\sim6$ quasars.
All these quasars were discovered by the Sloan Digital Sky Survey 
\citep[SDSS;][]{yor00} and were selected from Fan et al. 2000, 2001, and 2004 
(hereafter quasar discovery papers). They have extensive follow-up 
observations in X-ray \citep[e.g.][]{she06}, $Spitzer$ bands 
\citep[e.g.][]{jia06}, millimeter/submillimeter and radio wavelengths 
\citep[e.g.][]{wan07}, providing a fundamental dataset to study luminous 
quasars at $z\sim6$ in the context of the growth of early BHs and their 
relation to galaxy formation. The Gemini data have high SNRs and excellent 
NIR wavelength coverage. They are used to measure the metal abundances in
high-redshift environments and estimate central BH masses using empirical mass 
scaling relations \citep[e.g.][]{mcl04,ves06}.

We also use these high-redshift spectra to study the evolution of strong 
intergalactic \mgii\ absorption at $z<6$. Quasar absorption lines are an 
unbiased tracer of intervening gaseous material along the lines of sight to 
quasars. 
Strong \mgii\ absorption systems are usually directly associated with 
galaxy halos or disks \citep[e.g.][]{cha98,chu00,ste02}. The statistical
properties of the \mgiid\ doublet absorbers at low redshift ($z\le2.2$) have 
been well studied (e.g. Steidel \& Sargent 1992; Nestor et al. 2005; Prochter 
et al. 2006; see a summary of \mgii\ absorption surveys by Churchill
et al. 2005). At higher redshift, the \mgii\ doublet is shifted to the NIR 
wavelength range, where detectors are much less efficient than are detectors 
at the optical region. The study of high-redshift \mgii\ absorbers is 
therefore
limited by the small number of quasars with good NIR spectra available. Our 
Gemini/GNIRS spectra with high SNRs have excellent NIR wavelength coverage,
so they are efficient to probe the \mgii\ absorption systems at $z<6$.

The structure of the paper is as follows. In $\S$ 2 we describe our Gemini 
observations and data reduction of the six quasars. In $\S$ 3 we conduct
a detailed spectral analysis, including the measurement of emission-line 
strengths. Chemical abundances are calculated and central BH masses are 
estimated in $\S$ 4. \mgii\ absorption systems in the quasar spectra are 
identified and analyzed in $\S$ 5. We give a brief summary in $\S$ 6. 
In this paper we use $\lambda$ ($\lambda_0$) to denote 
observed-frame (rest-frame) wavelength. We use a $\Lambda$-dominated flat 
cosmology with H$_0$ = 70 km s$^{-1}$ Mpc$^{-1}$, $\Omega_{m}$ = 0.3 and
$\Omega_{\Lambda}$ = 0.7 \citep{spe07}.

\section{OBSERVATIONS AND DATA REDUCTION}

We obtained NIR spectra for the six luminous $z\sim6$ quasars using 
Gemini-South/GNIRS and Gemini-North/NIRI. Five of them were chosen to be 
easily reached by Gemini-South, and were observed using GNIRS in February and 
March 2006. The observations on GNIRS were carried out in cross-dispersion 
mode, giving a simultaneous wavelength coverage of 0.9--2.5 $\mu$m with 
excellent transmission. At $z\sim6$, strong diagnostic emission lines such as 
\civ\ and \mgii\ are redshifted to this wavelength range, so GNIRS provides us 
an efficient way to collect NIR information for high-redshift quasars. We used 
the short camera on GNIRS with a pixel scale of $0.15\arcsec$/pixel. The slit 
length in cross-dispersion mode is $6.1\arcsec$ and we used a slit width of 
$0.675\arcsec$ to optimize the combination of resolution and light gain. The 
resolving power is about 800 and the dispersion varies from $\sim$3 \AA\ per
pixel at 0.9 $\mu$m to $\sim$6 \AA\ per pixel at 2.5 $\mu$m. This is 
sufficient to separate the \mgiid\ doublet absorption lines. We used 
the standard ABBA nodding sequence between exposures. The exposure time at 
each nod position was five minutes and the distance between the two positions 
was $3\arcsec$.
Before or after the exposure of each quasar, a nearby A or F spectroscopic 
standard star was observed for flux calibration and to remove telluric
atmosphere absorption. The log of observations is given in Table 1, where
redshifts, $z_{AB}$ (AB magnitudes) and $J$ (Vega-based magnitudes) are taken 
from the quasar discovery papers.

The GNIRS data were reduced using the Gemini package within IRAF\footnote{IRAF 
is distributed by the National Optical Astronomy Observatories, which are 
operated by the Association of Universities for Research in Astronomy, Inc., 
under cooperative agreement with the National Science Foundation.}.
Briefly, for each quasar the NIR spectroscopic data were sky-subtracted for 
each pair of images taken at the two nod positions. Then the images were 
flat-fielded. In cross-dispersion mode, GNIRS uses orders 3--8 to cover 
0.9--2.5 $\mu$m. A single flat has non-uniform illumination between orders,
so three sets of flats with different exposure times were taken each night.
The final flat is extracted from the three flats so that each order has good 
illumination but without saturation. After flat-fielding, distortion 
correction and wavelength calibration were applied, the frames were 
combined, and one-dimensional spectra were extracted. Then the spectra
were flux calibrated and corrected for telluric atmosphere absorption using
the spectra of the spectroscopic standard stars. Finally the spectra in 
different orders were scaled and connected to a single spectrum. 
Adjacent orders cover a short common wavelength range, which enables the 
individual orders to be scaled to form a continuous, smoothly connected
spectrum. This arbitrarily scaled spectrum was then calibrated to the $J$ 
broad-band magnitude listed in Table 1 and thereby placed on an absolute flux 
scale with an uncertainty of 5$\sim$10\%. 
The typical SNR at 1.2 $\mu$m reaches $\sim8$ per pixel.

Figure 1 shows the spectra of the five quasars in the observed frame. The 
spectra at $\lambda\leq1.0$ $\mu$m are taken from the quasar discovery papers. 
The spectra have been smoothed by three pixels. The bottom panel shows the sky 
transparency at an airmass of 1.0 and water vapor of 1.6 mm. There are strong
telluric absorption bands around 1.4 and 1.9 $\mu$m, so we do not show
the spectra in these ranges. In Figure 1 we also show a spectrum of a standard 
star HIP 54027 calibrated using the same procedure as we do for our science 
objects. For comparison, a spectrum (in red) of a black body with the 
effective temperature of HIP 54027 is given on top of the spectrum of 
HIP 54027. The good agreement between the two spectra indicate that spectra
in different dispersion orders are properly scaled and connected.

We obtained a $K$-band spectrum for SDSS J1623+3112\footnote{The naming 
convention for SDSS sources is SDSS JHHMMSS.SS$\pm$DDMMSS.S, and the 
positions are expressed in J2000.0 coordinates as given in Table 1. We use 
SDSS JHHMM$\pm$DDMM for brevity.} using Gemini-North/NIRI in August and 
September 2004 and August 2005. The observations were made with a f/6 
camera in spectroscopic mode. The long-slit has a length of $110\arcsec$
and we chose to use a slit width of $0.75\arcsec$, which provides a resolving
power of $\sim500$. The observing strategy and data reduction are similar to
those used for the GNIRS data. The standard ABBA mode was used with a 
nodding distance of $20\arcsec$ and an exposure time of five minutes at each 
nod position. A nearby A or F spectroscopic standard star was observed before
or after the target observation. The NIRI data were also reduced using the 
Gemini package within IRAF.
After sky-subtraction, flat-fielding and wavelength calibration, the 
frames were combined. Then one-dimensional spectra were extracted from the 
combined images, and were flux calibrated using the spectra of standard stars.
The observing log is given in Table 1 and the spectrum is shown in Figure 2.
Since we do not have $K$-band photometry for this object, we cannot apply an
absolute flux calibration. In Figure 2, we have scaled the spectrum to 
its $J$ magnitude assuming a power-law continuum slope of $\alpha_{\nu}=-0.5$
($f_{\nu}\sim\nu^{\alpha_{\nu}}$).

\section{SPECTRAL FIT AND EMISSION-LINE MEASUREMENTS}

To analyze broad emission lines such as \civ\ and \mgii\ we fit and subtract 
the UV \feii\ emission that contaminates most of the UV-optical spectral 
region; thousands of \feii\ lines are emitted at UV-optical energies that 
blend with other lines in the spectrum. We use the UV \feii\ template 
presented by \citet{ves01} and adopt the fitting procedure described in detail 
in $\S$ 4.2 by these authors. Briefly, the fitting is done as an iteration 
over a power-law fit to the continuum emission and a model fit to the \feii\ 
emission using a scaled and broadened version of the \feii\ template. The
\feii\ template is broadened by convolving the template with Gaussians with
a range of sigmas. In the first iteration a power-law is fitted to regions 
with very little or no contribution from line emission (e.g. 1275--1295 \AA; 
1325--1335 \AA; 1425--1480 \AA; 2180--2220 \AA; 3000--3040 \AA). Upon 
subtraction of this primary continuum fit multiple broadened copies of the 
\feii\ template are scaled to regions which predominantly contain Fe
emission. The broadened, scaled \feii\ template that provides the best overall 
match to the observed \feii\ emission in the observed spectrum is then 
selected and subtracted. Some adjustments of the \feii\ fitting regions are 
sometimes necessary to optimize the fit, as judged by a minimization of the
residual flux upon subtraction of the \feii\ model. After subtraction of the 
best matching \feii\ model from the original data (i.e., with the continuum 
not subtracted) another power-law fit to the continuum emission is performed; 
the lack of (most of the) contaminating \feii\ enables the fitting to be done 
over a broader wavelength range, often resulting in a better and more 
realistic 
continuum fit. After subtraction of this continuum model, the \feii\ fitting 
is repeated. The iteration of the continuum and \feii\ fitting is repeated 
until these individual fits do not change and the subtraction of the \feii\ 
model leaves a smooth, featureless spectrum (barring the strong, broad non-Fe 
emission lines) with a realistic power-law slope. Convergence is often 
obtained in the second or third iteration. For SDSS J1623+3112, the wavelength 
range of the spectrum is too short to allow a reasonable fit to both the 
\feii\ emission and a power-law continuum, so we assume a slope of 
$\alpha_{\nu}=-0.5$ during the model fitting. This gives a lower limit of the 
\feii\ emission. Balmer continuum could be non-negligible at 
$\lambda_0\geq2000$ \AA\ \citep[e.g.][]{die02,die03b}, however, 
the wavelength coverage of our spectra is not long enough to provide a 
realistic fit to it, so we do not consider it in our fits. 
The power-law continuum fits to each of the quasars in our sample are listed 
in column 3 of Table 2. We note that the accuracy of the derived line width 
depends on the quality of the spectra. For widths broader than 5000--6000 km/s 
our ability to accurately determine the intrinsic \feii\ line width decreases.

We use \mgii\ emission lines to determine the redshifts of the quasars.
Accurate redshifts are important for various studies such as molecular line
searches. The redshifts given in the quasar discovery papers were mostly 
determined from the \lya\ emission lines, where strong \lya\ absorption 
systems usually cause large uncertainties in redshift measurements. While most 
of high-ionization, broad emission lines such as \civ\ can be significantly
blueshifted with respect to quasar systematic redshifts, the \mgii\ emission 
line has a small blueshift and thus provides a reliable redshift measurement
\citep[e.g.][]{mci99,van01,ric02}. The \mgii\ emission line at $z\sim6$ is 
close to a strong telluric absorption band at 1.9 $\mu$m, as seen in Figure 1.
To avoid the effect of the telluric absorption, we measure the line center of 
each \mgii\ emission line using a Gaussian profile to fit the top $\sim50$\% 
of the line. The line center is also estimated by calculating the 
$3\times$median--$2\times$mean mode of the top $\sim50$\% of the line. The two 
methods give similar results because the top $\sim50$\% of the line is almost 
symmetric. The measured redshifts are given in Column 2 of Table 2. They
are in good agreement with the values derived in \citet{kur07}, but differ by
0.02--0.03 from the values in the discovery papers. We did not detect the 
\mgii\ emission line for SDSS J0836+0054 and SDSS J1044$-$0125, which were 
observed with high humidity. In Table 2, the redshift of SDSS 
J0836+0054, measured also using the \mgii\ emission line, is taken from 
\citet{kur07}, and the redshift of SDSS J1044$-$0125 is measured from the 
\ciii\ $\lambda$1909 emission line using the same method.

We fit individual emission lines after the subtraction of the \feii\ emission
and power-law continuum. We use one Gaussian profile to fit most isolated 
lines and use multiple Gaussian profiles to fit blended lines. We allow both 
the line center and the line width to vary unless stated otherwise. Broad 
emission lines in quasars usually show asymmetric velocity profiles, so in the 
cases in which a single Gaussian profile does not provide a reasonable fit, we 
use double Gaussian profiles. There is no particular physical meaning to the
two Gaussian components. For each quasar, the following emission lines 
(if detected) are fitted.
(1) \lya\ + \nv\ + \siii\ $\lambda1262$ (hereafter \siii). We use four 
Gaussian profiles to fit the three lines, with the first two profiles
representing 
the broad and narrow components of \lya\ and the last two profiles
representing \nv\ and \siii, respectively. During the fit, the line centers 
are fixed at the values given by \citet{van01}. Since the blue side of the 
\lya\ line is strongly absorbed, we only fit the red side of the \lya\ 
line. (2) \oi\ $\lambda1304$, \cii\ $\lambda1335$, and \siiv\ $\lambda1397$
(hereafter \oi, \cii, and \siiv). We use one Gaussian profile to fit each of 
these lines. (3) \civ. The strong \civ\ line is fitted using one or two 
Gaussian components. (4) \heii\ + \oiii\ $\lambda1663$ (hereafter \oiii). 
These two lines are very weak, but \heii\ is important for determining 
chemical abundances. They were tentatively detected in three quasars and 
fitted simultaneously using two Gaussians. (5) \aliii\ $\lambda1857$ + \ciii\ 
$\lambda1909$ (hereafter \aliii\ and \ciii). The weak \aliii\ line is often 
detected on the blue wing of the strong \ciii\ line. We use two Gaussian 
profiles to fit these two lines simultaneously if \aliii\ is detected, 
otherwise the \ciii\ line is fitted using a single Gaussian profile. We ignore 
the weak \siiii\ $\lambda1892$ line on top of \ciii.
(6) \mgii. The \mgii\ emission line can be well fitted by double Gaussians
\citep[e.g.][]{bar03}, with a weak component located on the blue side of
a much stronger component. If the blue side of \mgii\ is affected by telluric 
absorptions, we fit the line using a single Gaussian. For the emission lines 
that are severely affected by broad-absorption lines (BALs), we fix their 
central wavelengths and fit their red half sides using a single Gaussian 
profile.

Figures 1 and 2 show the fitting results for the six quasars. The blue solid 
lines are the best fits to the emission lines and the blue dashed lines are
the best model fits to the power-law continuum and \feii\ emission. The red 
lines are the sums of all the components. For comparison, we show the 
low-redshift composite spectrum of \citet{van01} in gray. Table 3 shows the 
rest-frame EW and FWHM of emission lines derived from the model fitting 
results. For the lines which were fitted using double Gaussian profiles, their 
EW and FWHM are calculated from the combination of the two components. The 
errors given in the table are the formal uncertainties obtained from our 
fitting process, so the real errors could be larger in the cases in which the 
lines are heavily blended, or we fit a profile of the wrong shape.
The \heii\ emission is tentatively detected in
three quasars at a significance level of $\sim2\sigma$. We note from Table 2
that three quasars have continuum slopes consistent with the slope 
$\alpha_{\nu}=-0.44$ of the composite spectrum of low-redshift SDSS quasars 
\citep{van01}, while SDSS J1030+0524 and SDSS J1306+0356 show significantly
bluer continua with $\alpha_{\nu}\sim0.5$. \citet{pen03} obtained an average 
slope of $\alpha_{\nu}=-0.57$ with a scatter of 0.33 from 45 
intermediate-redshift SDSS quasars at $3.6<z<5.0$. This indicates that the 
continuum slopes in the two unusual quasars deviate from the average slopes
at a $\sim3\sigma$ significance level. Table 2 also shows the rest-frame 
optical continuum slopes derived from the $Spitzer$ IRAC broad-band photometry 
of the same objects \citep{jia06}. The optical continua are generally redder 
than the UV continua measured from the Gemini spectra, as also reported by 
\citet{van01}.

\section{CHEMICAL ABUNDANCES AND CENTRAL BH MASSES}

\subsection{Chemical Abundances}

Emission-line ratios can be used to measure gas metallicity in the quasar BLR
and track the chemical evolution with redshift. When weak emission lines are 
difficult to detect in high-redshift quasars, strong broad emission lines such 
as \nv, \civ, and \mgii\ become a powerful tool to probe metallicity at 
high redshift \citep[e.g.][]{ham92,ham93}. Different elements form on 
different timescales. Elements such as C, O, and Mg are 
formed in the explosions of massive stars, and their enrichment is rapid; 
while the second generation element N is produced from previously 
generated C and O, and its enrichment is relatively slow. Helium is 
a primordial element and its abundance changes little with cosmic time. 
Chemical abundances measured from emission lines and intrinsic absorption 
lines show solar or supersolar metallicity in quasar environments at low
redshift \citep[e.g.][]{ham07}, and can be as high as 15 times solar 
metallicity \citep{bal03}. Furthermore, studies have shown that there is 
little chemical evolution up to the highest known redshift 
\citep[e.g.][]{bar03}.

We calculate emission-line fluxes from the model fitting results described
in $\S$ 3. The fluxes are normalized to the \civ\ fluxes and shown in 
Table 4 and Figure 3
(filled circles with errors). \citet{nag06} used a sample of over 5000 quasars
from the SDSS Data Release Two to study quasar BLRs. They made quasar composite 
spectra in the ranges of $2.0\le z \le4.5$ and $-29.5\le M_B \le-24.5$, and 
measured emission-line ratios and metallicities in the composite spectra for 
each redshift and luminosity bin. They found that most of the line ratios 
(with respect to the \civ\ line) that they investigated do not show strong 
evolution with redshift. For the purpose of comparison, we show in Figure 3 
the line ratios of quasars (filled squares) in the luminosity range of
$-27.5<M_B<-26.5$ from \citet{nag06}. Although there are correlations between 
quasar luminosities and most line ratios \citep{nag06}, most of our quasars 
have luminosities in a small range of $-27.5<M_B<-26.5$. Figure 3 shows 
that most of the flux ratios do not exhibit strong evolution up to $z\sim6$. 
There is a trend for higher \cii/\civ\ and \oi/\civ\ ratios at higher 
redshift. \citet{nag06} already noted that the \oi/\civ\ flux ratio is 
marginally correlated with redshift. This trend could also be caused by 
the intrinsic dispersion of the flux ratios.

Photoionization models have shown that a series of emission-line ratios can
be used to estimate gas metallicity in the BLRs of quasars. \citet{ham02} 
studied the relative N abundance as a metallicity indicator based on 
the locally optimally emitting cloud (LOC) model \citep{bal95}, and calculated 
theoretical emission-line ratios as a function of metallicity Z/Z$_\sun$. We 
estimate the metallicity from the \nv/\civ\ and \nv/\heii\ ratios using Figure 
5 of \citet{ham02}. There is a 30\% difference between the solar abundances
used by \citet{ham02} and the latest solar abundances \citep{bal03}, so we
scale the abundances of \citet{ham02} to match the latest values using the
method by \citet{bal03}. The results are shown as the open circles 
in Figure 4.
\citet{nag06} also used the LOC model, but included more UV emission lines in 
their photoionization model. We use Figure 29 of \citet{nag06} to measure the 
metallicity from various line ratios and show the measurements as the filled 
circles in Figure 4. The two models give similar metallicities from the 
\nv/\civ\ and \nv/\heii\ ratios. The metallicities in these high-redshift 
quasars are supersolar, with a typical value of $\sim4$ Z$_{\sun}$.
\citet{nag06} computed chemical abundances as a function of redshift for the 
SDSS composite spectra using their photoionization model. The filled squares 
in Figure 4 show the abundances averaged in the luminosity range of 
$-28.5<M_B<-25.5$. The filled triangles represent the chemical abundances of 
eleven $3.9<z<5.0$ quasars from \citet{die03a}. Figure 4 shows that the
metallicity is consistent with no strong evolution within the errors up to 
$z\sim6$.

The \feii/\mgii\ ratio is important in understanding chemical evolution at 
high redshift. We use the \feii\ emission line complex at 
$\lambda_0=2000\sim3000$ \AA\ as the \feii\ flux indicator 
\citep[e.g.][]{die02,bar03}. The flux of the \feii\ complex is integrated from 
2200 to 3090 \AA\ over the best-fitting \feii\ template. The derived 
\feii/\mgii\ ratios (filled circles) with a typical value of $4.9\pm1.4$ are 
shown in Table 4 and Figure 5. Using a similar method \citet{kur07} 
obtained a lower ratio of $2.7\pm0.8$ from their quasar sample.
The discrepancy is partly due to the fact that they corrected for Balmer 
continuum in their spectral fits. For 
comparison, we also show the \feii/\mgii\ ratios of $z\sim6$ quasars from 
previous measurements \citep{bar03,mai03,fre03,iwa04} as open circles in 
Figure 5. \citet{iwa02} measured the
\feii/\mgii\ ratios from the spectra of quasars at $0<z<5.3$, and computed
the median values of the ratios for a range of redshifts. These median
ratios are shown as the filled squares in Figure 5. To study the evolution
of \feii/\mgii, \citet{die03b} made quasar composite spectra from $z=0$ to
5. The \feii/\mgii\ ratios measured from the composite spectra are shown
as the filled triangles in Figure 5. We note that the \feii\ flux in some
studies was integrated over the wavelength range of $2150<\lambda_0<3300$ \AA\
\citep[e.g.][]{fre03,iwa04}, slightly different from the range that we used. 
We also note that some of these studies used different \feii\ templates to 
measure \feii\ fluxes. These issues usually affect the \feii/\mgii\ 
measurements by less than a factor of two and cause a relatively large scatter 
in the measurements at $z\sim6$. Figure 5 shows that our results are 
consistent with those derived from both low-redshift samples and other 
$z\sim6$ samples within errors. Because the details of how the \feii\ bump is 
formed are not well understood \citep[e.g.][]{bal04}, so we do not derive 
actual abundances from the \feii/\mgii\ ratios.

We have shown that the metallicity in the BLRs of high-redshift quasars is
supersolar, and the lack of strong evolution in metallicity continues to 
$z\sim6$. The high 
metallicity at $z\sim6$ indicates that vigorous star formation and element 
enrichment have occurred in quasar host galaxies in the first Gyr of 
cosmic time. Millimeter and submillimeter observations revealed that
luminous $z\sim6$ quasars are extremely far-infrared (FIR) luminous 
($\sim10^{13}$ L$_{\sun}$) and have a large amount of cool dust 
($10^8$--$10^9$
M$_{\sun}$) \citep{car01,ber03a,pri03,rob04}. If the FIR emission
is mainly powered by star formation in host galaxies, the star formation rates
are estimated to be $\sim1000$ M$_{\sun}$ yr$^{-1}$. Strong starbursts can be 
induced by merging gas-rich galaxies \citep[e.g.][]{hop06,li07}.
CO observations have already revealed the presence of 
$\sim2\times 10^{10}$ M$_{\sun}$ of molecular gas in the highest redshift 
quasar SDSS J1148+5251 \citep{wal03,ber03b,wal04}. This amount of gas can 
sustain the star formation rate of $\sim3000$ M$_{\sun}$ yr$^{-1}$, inferred
from the FIR luminosity, for $10^7$--$10^8$ years. If the typical duty cycle 
time of luminous quasars is $10^7$--$10^8$ years 
\citep[e.g.][]{kau00,wyi03,she07}, it would have enough time to form a large 
amount of first generation elements from massive stars.

We have also shown that the \feii/\mgii\ ratios of the $z\sim6$ quasars
are consistent with low-redshift measurements. Most of the Fe in the solar
neighborhood is produced in intermediate-mass stars, and their enrichment is 
delayed by $\sim1$ Gyr compared to Mg.
Mg$^+$ and Fe$^+$ have similar ionization potentials and thus the \feii/\mgii\
ratio reflects the Fe/Mg abundance \citep{die03b}. Therefore, if Fe was 
generated in the same way as it is in our neighbors, the \feii/\mgii\ ratio is 
expected to be a strong function of cosmic time at high redshift.
However, we do not see any evolution in this ratio even at $z\sim6$. 
It has been shown that the timescale of Fe enrichment from SNe Ia strongly
depends on star formation histories of host galaxies, and can be much shorter
in high-redshift quasar environments \citep[e.g.][]{fri98,mat01}. 
\citet{mat01} found that the timescale for the maximum SN Ia rate (Fe
enrichment) can be as short as $\sim0.3$ Gyr in an elliptical galaxy with a 
high star formation rate but a short star formation history. On the other 
hand, \citet{ven04} pointed out that stars with a present-day initial
mass function are sufficient to produce the observed \feii/\mgii\ ratios in
$z\sim6$ quasars, and SNe Ia are not necessarily the main contributor.
Fe could also be generated in Population III stars, which are suggested to 
be first generation stars with masses between 100 and 1000 M$_{\sun}$. These 
very massive stars produce a large amount of 
Fe within a few Myr due to their short lifetimes \citep{heg02}. 

\subsection{Central BH Masses}

The central BH masses of high-redshift quasars are important in understanding
the growth of BHs and quasar accretion rates in the early universe. For 
high-redshift quasars where direct mass measurements are difficult, BH masses 
can be estimated using mass scaling relations based on broad emission line 
widths and continuum luminosities. Strong emission lines, including \hb, \civ, 
and \mgii, have been used to determine BH masses 
\citep[e.g.][]{wan99,die04,mcl04,ves06}. Based on empirical scaling relations, 
a few luminous $z\sim6$ quasars have been found to harbor central BHs with 
masses of $10^9\sim10^{10}$ M$_{\sun}$ \citep[e.g.][]{bar03,ves04}.

We estimate the central BH masses for the six quasars using the following BH
mass scaling relations,
\begin{equation}
  \rm M_{BH}(\mbox{\civ}) = 4.57 \left( \frac{FWHM(\mbox{\civ})}{km\ s^{-1}}
  \right)^2 \left(\frac{\lambda L_{\lambda}(1350\mbox{\AA})}
  {10^{44}\ ergs\ s^{-1}}\right)^{0.53} M_{\sun}
\end{equation}
and
\begin{equation}
  \rm M_{BH}(\mbox{\mgii}) = 3.2 \left( \frac{FWHM(\mbox{\mgii})}{km\ s^{-1}}
  \right)^2 \left(\frac{\lambda L_{\lambda}(3000\mbox{\AA})}
  {10^{44}\ ergs\ s^{-1}}\right)^{0.62} M_{\sun}
  \label{eq:mgii}
\end{equation}
derived by \citet{ves06} and \citet{mcl04}, respectively.
$\rm L_{\lambda}(1350\mbox{\AA})$ and $\rm L_{\lambda}(3000\mbox{\AA})$ are
luminosities at rest-frame 1350 and 3000 \AA. The results are listed in 
Columns 4 and 5 of Table 5. They are in agreement with the masses derived by
\citet{kur07}. For SDSS J1030+0524 and SDSS J1306+0356, the \civ--based BH 
masses are greater than the \mgii--based BH masses by a factor of $\sim$3, but 
still within the intrinsic scatter (a factor of 3--4) in these scaling 
relations. The unusual blue continua in the two quasars could cause this 
discrepancy. The \mgii--based BH masses are also estimated using the latest
scaling relation based on FWHM(\mgii) and $\rm L_{\lambda}(1350\mbox{\AA})$
by Vestergaard et al. (in preparation). 
The results are shown in Column 6, and 
they are consistent with the \civ--based BH masses.  Similar to other luminous 
quasars at $z\sim6$, our quasars have central BH masses of 
$10^9$--$10^{10}$ M$_{\sun}$ \citep[e.g.][]{bar03,ves04,jia06}. 

We calculate the Eddington luminosity ratios of the quasars using their
bolometric luminosities and BH masses. We use the \mgii--based BH mass 
for SDSS J1623+3112 and the \civ--based BH masses for the others. The 
bolometric luminosities were calculated from multiband observations of the
quasars by \citet{jia06}. The results
are given in Column 3 of Table 5. These luminous $z\sim6$ quasars have 
Eddington ratios of order unity, comparable to quasars with similar 
luminosities at lower redshift \citep[e.g.][]{mcl04,ves04,kol06}.

It is remarkable that billion-solar-mass BHs can form less than one Gyr after 
the Big Bang. With the reasonable assumption that BH accretion is at the 
Eddington rate, the BH mass at time $t$ is $M_t=M_0e^{t/\epsilon\tau}$,
where $\epsilon$ is the radiative efficiency, $\tau=4.5\times10^8$ years, and 
$M_0$ is the initial BH mass or the seed BH mass. Seed BHs can be produced 
from the collapse of Population III stars or gas clouds, and their masses are
roughly $10^2$--$10^4$ M$_{\sun}$ \citep[e.g.][]{mad01,vol06,lod07}. Consider
the case in which a massive BH at $z=6$ formed from a seed BH with $M_0=10^3$ 
M$_{\sun}$ at $z=20$. The $e$-folding time $\epsilon\tau$ for the BH growth is 
roughly $4.5\times10^7$ years if $\epsilon\sim0.1$. The BH grows from $z=20$ 
to 6 by 15 $e$-foldings, or a factor of $3.3\times10^6$, which results in a 
massive BH with $M_t=3.3\times10^9$ M$_{\sun}$ at $z=6$, comparable to the 
observed BH masses in our sample. If a quasar is shining at half of the 
Eddington limit, its BH grows from $z=20$ to 6 by only 7.5 $e$-foldings, or 
a factor of $\sim2000$, making it very difficult to form billion-solar-mass 
BHs in this scenario. In addition, if Eddington-limited accretion is via 
standard thin disks, BHs are likely to be spun up and the radiative efficiency 
and 
Eddington timescale will increase \citep{vol06,ree06}. In this case it would
take much longer to form massive BHs. So super-Eddington accretion or lower 
radiative efficiency is probably required to form BHs with 
$M_t=10^9$--$10^{10}$ M$_{\sun}$ by $z=6$.

\subsection{Notes on Individual Objects}

{\bf SDSS J0836+0054 ($z=5.810$).}
SDSS J0836+0054 was discovered by \citet{fan01}. Its redshift estimated from 
the \lya\ emission line is 5.82,
comparable to the redshift $5.810\pm0.003$ determined from the \mgii\ emission
line \citep{kur07}. Due to high humidity, we did not detect the \mgii\ 
emission line. \citet{ste03} found a very red continuum slope of
$\alpha_{\nu}=-1.55$ in their NIR spectrum of this quasar. However, using the
GNIRS spectra with a longer wavelength coverage and a higher spectral quality
we obtained a slope of $\alpha_{\nu}=-0.62$, which is close to
$\alpha_{\nu}=-0.44$, the slope of the SDSS composite spectrum \citep{van01}. 
SDSS J0836+0054 is the most luminous
quasar known at $z>5.7$. The central BH mass estimated from the \civ\ emission
line is $9.5\times10^9$ M$_{\sun}$. It was also detected by the
Faint Images of the Radio Sky at Twenty-cm \citep[FIRST;][]{bec95}.
The flux at 1.4 GHz measured by FIRST is 1.11$\pm$0.15 mJy and measured by
\citet{pet03} is 1.75$\pm$0.04 mJy.

{\bf SDSS J1030+0524 ($z=6.309$).}
SDSS J1030+0524 is the second highest redshift quasar known to date. The
redshift estimated from \lya\ is 6.28 \citep{fan01}. We obtained a more
accurate and slightly higher redshift $6.309\pm0.009$ from \mgii, consistent
with the value 6.311 of \citet{iwa04}. The emission-line ratios and
metallicity in this object have been studied by a few groups. In the discovery
paper \citet{fan01} estimated lower limits for the \nv/\civ\ and \nv/\heii\
ratios: \nv/\civ\ $\ge0.4$ and \nv/\heii\ $\ge3.0$. Using VLT NIR observations
\citet{pen02} found \nv/\civ\ $=0.35$ and \nv/\heii\ $>4.3$. From our
measurements we obtained \nv/\civ\ $=0.21\pm0.02$ and \nv/\heii\
$=2.85\pm1.70$, comparable to the previous studies. The \feii/\mgii\ ratios
of this object measured in previous papers are quite different. The values
reported by \citet{mai03}, \citet{fre03}, and \citet{iwa04} are $8.65\pm2.47$,
$2.1\pm1.1$, and $0.99_{-0.99}^{+1.86}$, respectively. We obtained
\feii/\mgii\ $=5.46\pm0.90$, consistent with the result of \citet{mai03}. Note 
that the \feii\ flux in \citet{fre03} and \citet{iwa04} was integrated over 
the wavelength range $2150<\lambda_0<3300$ \AA, slightly different from the 
range $2200<\lambda_0<3090$ \AA\ that \citet{mai03} and we used. In addition,
our NIR spectra have much higher SNRs than those of previous studies.

{\bf SDSS J1044$-$0125 ($z=5.778$).}
SDSS J1044$-$0125 was the first quasar discovered at $z>5.7$ \citep{fan00}.
The redshift estimated from \lya\ is 5.80 \citep{fan00} and that estimated 
from \civ\ is 5.74 \citep{goo01}. However, these measurements could be 
severely affected by absorption since this is a BAL quasar 
\citep{goo01,djo01}. Due to high humidity we 
did not detect its \mgii\ emission line. The redshift measured from \ciii\ 
(Figure 1 shows that this line is not affected by absorption) is 
$5.778\pm0.005$, consistent with the value 5.78 obtained by \citet{fre03}. 
\citet{fre03} also reported a \feii/\mgii\ ratio of $5.0\pm2.1$ for this 
quasar. SDSS J1044$-$0125 has the highest BH mass of the objects in our 
sample. Submillimeter observations \citep{pri03} and $Spitzer$
observations \citep{jia06} revealed a large amount of cool and hot dust
in this object. SDSS J1044$-$0125 is the only BAL quasar in our sample.
The fraction of BAL quasars in the sample is 16.7\%, similar to the 
low-redshift fraction \citep{tru06}.

{\bf SDSS J1306+0356 ($z=6.016$).}
The discovery paper for SDSS J1306+0356 indicated a redshift $z=5.99$ 
based on \lya\ \citep{fan01}.
We obtained a more accurate and slightly higher redshift $6.016\pm0.005$.
\citet{pen02} reported a \nv/\civ\ ratio of 0.67 for this quasar. The ratio
we measured is $0.55\pm0.05$. We also obtained a \feii/\mgii\ ratio of 
$5.77\pm0.72$, slightly smaller than the value $9.03\pm2.26$ derived by 
\citet{mai03} based on low SNR spectra. We detected four strong \mgiid\ 
absorption systems in the spectra of this quasar, including the two highest 
redshift \mgii\ absorbers known (see $\S$ 5).

{\bf SDSS J1411+1217 ($z=5.927$).}
SDSS J1411+1217 has the narrowest emission lines in our sample. Its \lya\
and \nv\ emission lines are well separated. Therefore the redshift $z=5.93$
measured from \nv\ and \oi\ in the discovery paper \citep{fan04} is accurate,
and is consistent with the redshift $5.927\pm0.004$ measured from \mgii\
in the Gemini spectrum. SDSS J1411+1217 has the lowest BH mass in the sample.
It also has a very weak $Spitzer$ 24$\mu$m ($\lambda_0 \sim 3.5$ $\mu$m) flux,
and its SED does not show any hot dust emission \citep{jia06}. Another quasar
SDSS J0005$-$0006 (not in this sample) shows similar IR properties and does 
not have hot dust emission either. This may reflect dust evolution at high 
redshift, as such dustless quasars are unknown at lower redshift.

{\bf SDSS J1623+3112 ($z=6.247$).}
The redshift of SDSS J1623+3112 measured from \lya\ in the discovery paper
is 6.22. We obtained a higher redshift of $6.247\pm0.005$ from \mgii.
This makes it the third highest redshift quasar known to date.

\section{\mgii\ ABSORPTION SYSTEMS AT $z<6$}

We search for \mgiid\ intervening systems in the spectra of five $z\sim6$ 
quasars. SDSS J1623+3112 is excluded in this analysis because it has a short 
wavelength coverage. The SNRs of the spectra for different quasars are 
different, as are the dispersions from one order to another. The SNRs at 
$\lambda\le 1.2$ $\mu$m are usually higher than those at longer wavelengths, 
especially in the range of $1.4<\lambda<1.8$ $\mu$m (roughly $H$ band). 
However, a \mgii\ $\lambda2796$ absorption line with rest equivalent width
$W_0^{\lambda2796}$ greater than 1.0 \AA\ at $z<3$ or a \mgii\ $\lambda2796$ 
absorption line with $W_0^{\lambda2796}$ greater than 1.5 \AA\ at $z>3$ can be 
detected at a $\sim7\sigma$ significance level in the lowest SNR parts of the 
spectra shown in Figure 1. 
Moreover, previous studies have used similar limits \citep[e.g.][]{nes05}. Our 
sample is small, and does not provide good statistics by itself, so the 
comparison with other samples is important. Thus to achieve unbiased 
statistics, we search for \mgii\ absorbers with $W_0^{\lambda2796}>1.0$ \AA\ 
at $z<3$ and \mgii\ absorbers with $W_0^{\lambda2796}>1.5$ \AA\ at $z>3$.

We search for pairs of absorption lines with separation matching the \mgiid\
doublet. Then we search for absorption lines of other species associated 
with the identified \mgii\ absorbers by matching various strong quasar 
absorption lines known in the literature to the spectra. Five \mgii\ 
absorption systems are identified, with one in the spectrum of SDSS J0836+0054 
and the other four in the spectra of SDSS J1306+0356.
We normalize the absorption lines using the local continua measured
by fitting low-order spline curves to the spectra. Figure 6 shows the 
normalized spectra of the five absorption systems, with the identified lines 
marked. These lines are usually strong and commonly detected 
in quasar absorption systems, such as \civd\ (marked as \civ-D in the figure),
\feiiaa, \aliia, \aliiid\ (marked as \aliii-D), \feiiab, \mgiid\
(marked as \mgii-D), and \mgia. The four absorption systems found in the
spectra of SDSS J1306+0356 are denoted as (a), (b), (c), and (d).

We measure the redshifts ($z_{abs}$) and rest-frame equivalent widths ($W_0$) 
of the absorption lines by fitting a Gaussian to each line. Unresolved or 
very weak absorption lines are not fitted. The results are given in Table 6. 
The \mgii\ doublet ratios are close to one, indicating that these strong 
\mgii\ absorbers are saturated. The \mgii\ absorber identified in the spectra 
of SDSS J0836+0054 is at $z=3.742$. A pair of moderate strong \aliiid\ lines 
in the same absorber are also detected. \civd, which is usually associated 
with \mgii, is beyond the spectral coverage. We do not detect any strong 
\feii\ absorption in this system. There are four \mgii\ absorption systems 
identified in the spectra of SDSS J1306+0356. Systems (a) and (b) are at 
relatively low redshifts. System (a) at $z=2.202$ has a strong \mgia\ 
absorption line. System (b) at $z=2.532$ has a strong \feiiab\ absorption 
line. \civd\ in both systems is beyond the spectral coverage. We discovered
two strong \mgiid\ absorbers at $z=4.8668$ and 
4.8823. They are the most distant \mgii\ absorbers known to date. The \mgii\ 
$\lambda2803$ line in system (c) and the \mgii\ $\lambda2796$ line in system 
(d) almost exactly overlap with each other, so we do not calculate $W_0$ for 
the two lines. Note that this is not the so-called line-locking effect, which 
mainly occurs in BAL quasars or other associated absorption systems with 
$z_{abs}\sim z_{em}$ \citep[e.g.][]{fol87,vil01,fec04}, since the two 
absorbers are far from SDSS J1306+0356. The proper distance between the two 
absorbers is about 1.4 Mpc, so they are not likely associated with the same 
galaxy, but could still belong to the same large-scale structure. Both 
absorption systems have strong \aliia\ and \feiiab\ absorption and weak 
\feiiaa\ and \aliiid\ absorptions. The \civd\ absorption is 
strong in system (c) but very weak in system (d). It has been found that 
\civd\ is almost always detected in strong \mgii\ absorption systems at low 
redshift \citep[e.g.][]{ste92}. The weakness of \civ\ in system (d) indicates 
that it is a rare, \civ-deficient system \citep{chu00}. \citet{els96} 
discovered a strong \mgii\ absorber at $z=4.38$ in the spectrum of a $z=4.68$
quasar. They found that the \civ\ absorption in this system is also very weak. 
Thus among the three strong \mgii\ absorbers at $z>4$, two have very weak 
associated \civ, probably indicating that $z>4$ absorption systems are in a
lower ionization state than their low-redshift counterparts \citep{els96}.

We calculate the comoving line-of-sight number densities ($dN/dz$) of the 
\mgii\ absorbers. The sample of the five \mgii\ absorbers is divided into two 
subsamples, one with $W_0^{\lambda2796}>1$ \AA\ at $z<3$ and the other one 
with $W_0^{\lambda2796}>1.5$ \AA\ at $z>3$. The redshift path covered by a 
spectrum is,
\begin{equation}
\Delta Z(W_0^{\lambda2796}) = \int_{\Delta z} g(W_0^{\lambda2796},z) dz,
\end{equation}
where $g(W_0^{\lambda2796},z)=1$ if a \mgii\ absorber at redshift $z$ 
and of the given EW would be detected in the spectrum in the redshift range 
$\Delta z$ and 
$g(W_0^{\lambda2796},z)=0$ otherwise. The total redshift paths covered by
the two subsamples are 3.4 and 10.4, respectively. Then the number density 
along the redshift path for each subsample is estimated by,
\begin{equation}
dN/dz = \frac{N_a} {\sum\ \Delta Z(W_0^{\lambda2796})_i},
\end{equation}
where $N_a$ is the number of \mgii\ absorbers and the sum is over all spectra 
in the subsample. We show our results in Figure 7. The filled
diamond represents the density for $W_0^{\lambda2796}>1.0$ \AA\ at $z<3$ and 
the filled circle represents the density for $W_0^{\lambda2796}>1.5$ \AA\ 
at $z>3$. The two data points are positioned at the median redshifts $z=2.6$ 
and 4.3, respectively. The statistical uncertainties are Poisson.

The number density of \mgii\ absorbers at low redshift has been well studied. 
\citet{ste92} collected the spectra of 103 quasars and found that the density 
of \mgii\ absorbers with $W_0^{\lambda2796}>0.3$ \AA\ does not shows cosmic
evolution, but the density of strongest \mgii\ absorbers significantly 
increases with redshift. \citet{nes05} reached similar results by studying 
\mgii\ absorptions in the spectra of thousands of quasars from the SDSS Early 
Data Release \citep{sto02}. They divided the sample into subsamples with 
$W_0^{\lambda2796}>$ 0.3, 0.6, 1.0, 1.5, 2.0, 2.5, and 3.0 \AA, respectively, 
and concluded that $dN/dz$ of the $W_0^{\lambda2796}>$ 0.3, 0.6, 1.0, and 
1.5 \AA\ subsamples are consistent with no evolution within errors, while the 
densities of the other three subsamples show strong evolution with redshift. 
For comparison, we show in Figure 7 the $W_0^{\lambda2796}>$ 1.0 and 1.5 \AA\ 
samples (open diamonds and circles) of \citet{nes05}. \citet{pro06} studied 
strong \mgii\ absorption systems using an even larger quasar sample from the 
SDSS Data Release Three \citep{aba04}. They found that there is no evolution
in the densities of $W_0^{\lambda2796}>$ 1.0 \AA\ absorbers at $0.8<z<2.0$. 
The crosses in Figure 7 show the densities 
for their $W_0^{\lambda2796}>1.0$ \AA\ sample 
(they did not use a sample of $W_0^{\lambda2796}>1.5$ \AA). In Figure 7 we 
also show the non-evolution curves (dashed lines) for the cosmology of 
$(\Omega_m, \Omega_{\Lambda}, h)=(0.3, 0.7, 0.7)$ with zero curvature,
\begin{equation}
dN/dz = N_0 (1+z)^2 [\Omega_m(1+z)^3 + \Omega_{\Lambda}]^{-1/2},
\end{equation}
where $N_0$ is a normalization constant. The curves have been scaled to 
minimize the $\chi^2$ to the data of \citet{nes05}. The density of \mgii\ 
absorbers with $W_0^{\lambda2796}>1.5$ \AA\ that we have measured is 
consistent with no cosmic evolution up to $z>4$. However, due to the large 
uncertainty, we cannot rule out the density evolution of \mgii\ absorbers.

\section{SUMMARY}

We have obtained high-quality NIR spectra of six luminous $z\sim6$ SDSS 
quasars from our Gemini observations. Five of them were observed using 
Gemini-South/GNIRS in cross dispersion mode, which provides a simultaneous 
coverage of 0.9--2.5 $\mu$m. The sixth was observed in $K$ band with 
Gemini-North/NIRI. These spectra have higher SNRs and better wavelength 
coverage than those in previous studies. We use the NIR spectra 
combined with optical data to study chemical abundances and BH masses in 
the $z\sim6$ quasars and strong intergalactic \mgii\ absorption at $2.2<z<6$. 

The spectra are fitted using a combination of a power-law continuum, an \feii\ 
emission template, and a series of emission lines. We calculate the fluxes for 
detected emission lines based on the Gaussian model fitting results. We find 
that the 
line flux ratios (normalized to the \civ\ fluxes) that we have investigated 
do not show strong evolution with redshift. We calculate gas metallicity
from emission-line ratios using the results of photoionization models given by 
previous studies and find that the metallicity in the BLRs of these 
high-redshift quasars is supersolar with a typical value of $\sim4$ 
Z$_{\sun}$. The comparison with low-redshift observations shows no show strong
evolution in metallicity up to $z\sim6$. 
The \feii/\mgii\ ratio is also measured for each quasar. We find a typical 
\feii/\mgii\ ratio of $4.9\pm1.4$, which is consistent with 
low-redshift samples. All these measurement indicate the existence of vigorous 
star formation and element enrichment in host galaxies in the first Gyr after 
the Big Bang. We have estimated central BH masses from the \civ\ and \mgii\ 
emission lines using empirical mass scaling relations. As found in other 
luminous $z\sim6$ quasars, these quasars have BH masses of $10^9$--$10^{10}$ 
M$_{\sun}$ and Eddington luminosity ratios of order unity. 

We have searched for strong \mgiid\ intervening systems in the spectra of five
quasars. Two \mgii\ absorbers with rest equivalent width 
$W_0^{\lambda2796}>1$ \AA\ at $2.2<z<3$
and three absorbers with $W_0^{\lambda2796}>1.5$ \AA\ at $z>3$ are identified
in the spectra of two quasars. The two most distant absorbers are at 
$z=4.8668$ and 4.8823, respectively. These are the highest redshift \mgii\ 
absorbers known to date. We calculate the comoving line-of-sight
number densities for the five identified \mgii\ absorbers. By comparing with
low-redshift studies we find that the densities ($dN/dz$) of \mgii\ absorbers
with $W_0^{\lambda2796}>1.5$ \AA\ are consistent with no cosmic evolution up 
to $z>4$. We note that our sample is small. A larger sample is needed to 
provide a good constraint on the density evolution of \mgii\ absorbers.

\acknowledgments

We thank the Gemini staff for their expert help in preparing and carrying out 
the observations. We thank T. Nagao and D. B. Nestor for providing useful 
data. We acknowledge support from NSF grant AST-0307384, a Sloan Research 
Fellowship and a Packard Fellowship for Science and Engineering 
(LJ, XF, MV, BCK). We acknowledge support from DFG grant SFB 439 (JDK).

Funding for the SDSS and SDSS-II has been provided by the Alfred P. Sloan
Foundation, the Participating Institutions, the National Science Foundation,
the U.S. Department of Energy, the National Aeronautics and Space
Administration, the Japanese Monbukagakusho, the Max Planck Society, and the
Higher Education Funding Council for England. The SDSS is managed by the 
Astrophysical Research Consortium for the Participating Institutions.
The SDSS Web Site is http://www.sdss.org/.

\clearpage
\begin{figure}
\epsscale{0.8}
\plotone{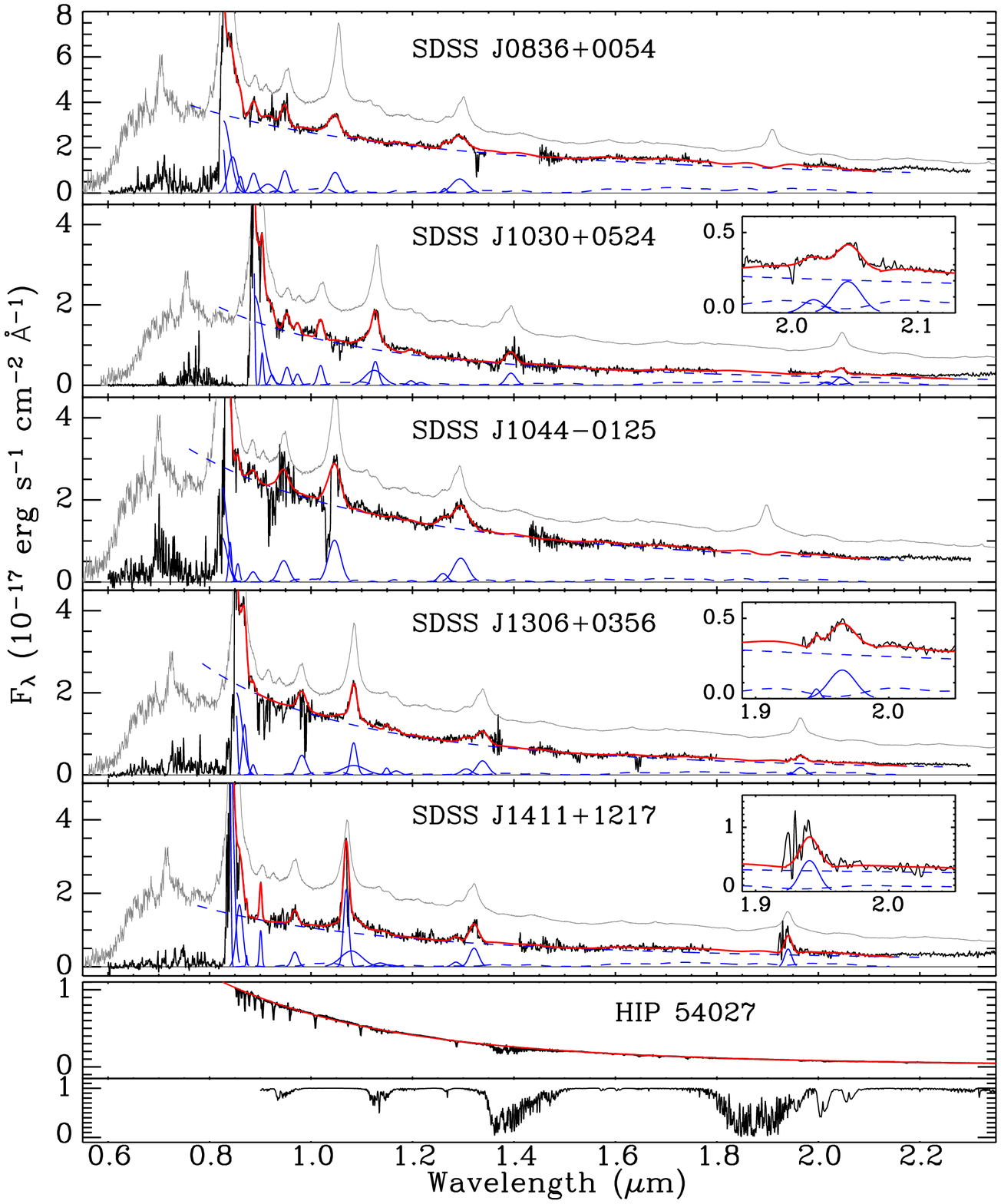}
\caption{Optical and NIR spectra of five $z\sim6$ quasars in our sample. 
The spectra at $\lambda>1.0$ $\mu$m were obtained with Gemini-South/GNIRS and
the spectra at $\lambda<1.0$ $\mu$m are taken from the quasar discovery 
papers. The spectra have been smoothed by three pixels.
The blue solid lines are the best fits to emission lines and the blue 
dashed lines are the best fits to the power-law continuum and \feii\ emission.
The red lines are the sums of all the components. For comparison, we show
the low-redshift composite spectrum of \citet{van01} in gray. 
A spectrum of a standard star HIP 54027 is also shown in the figure with 
a black body (with the effective temperature of HIP 54027) spectrum (in red) 
on top of it. The bottom panel 
shows the sky transparency at an airmass of 1.0 and water vapor of 1.6 mm.}
\end{figure}

\clearpage
\begin{figure}
\epsscale{0.6}
\plotone{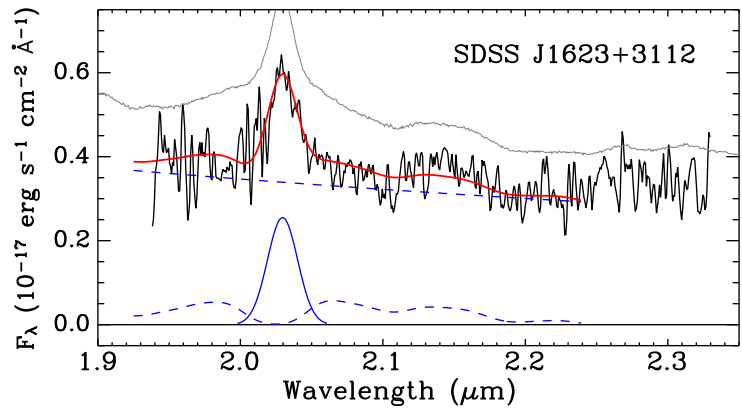}
\caption{Same as Fig. 1 for SDSS J1623+3112, which was observed in $K$ band 
using Gemini-North/NIRI.}
\end{figure}

\clearpage
\begin{figure}
\epsscale{1.0}
\plotone{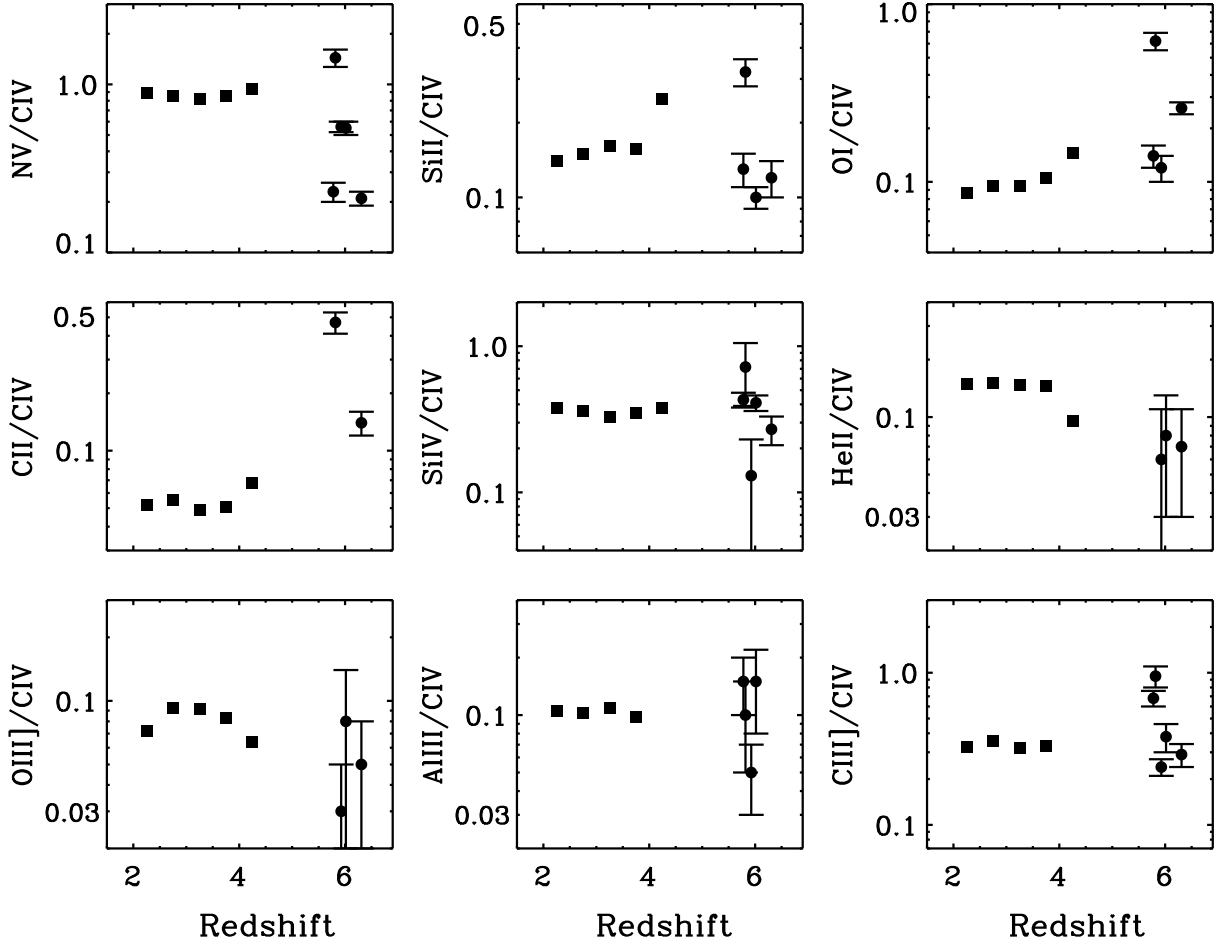}
\caption{Emission line fluxes (filled circles with errors) of the $z\sim6$
quasars compared to low-redshift measurements. The fluxes have been normalized 
to the \civ\ fluxes. The filled squares represent the flux ratios measured in 
the composite spectra of quasars in the luminosity range $-27.5<M_B<-26.5$ 
from \citet{nag06}.
Most of the flux ratios do not exhibit strong evolution up to $z\sim6$.}
\end{figure}

\clearpage
\begin{figure}
\epsscale{0.6}
\plotone{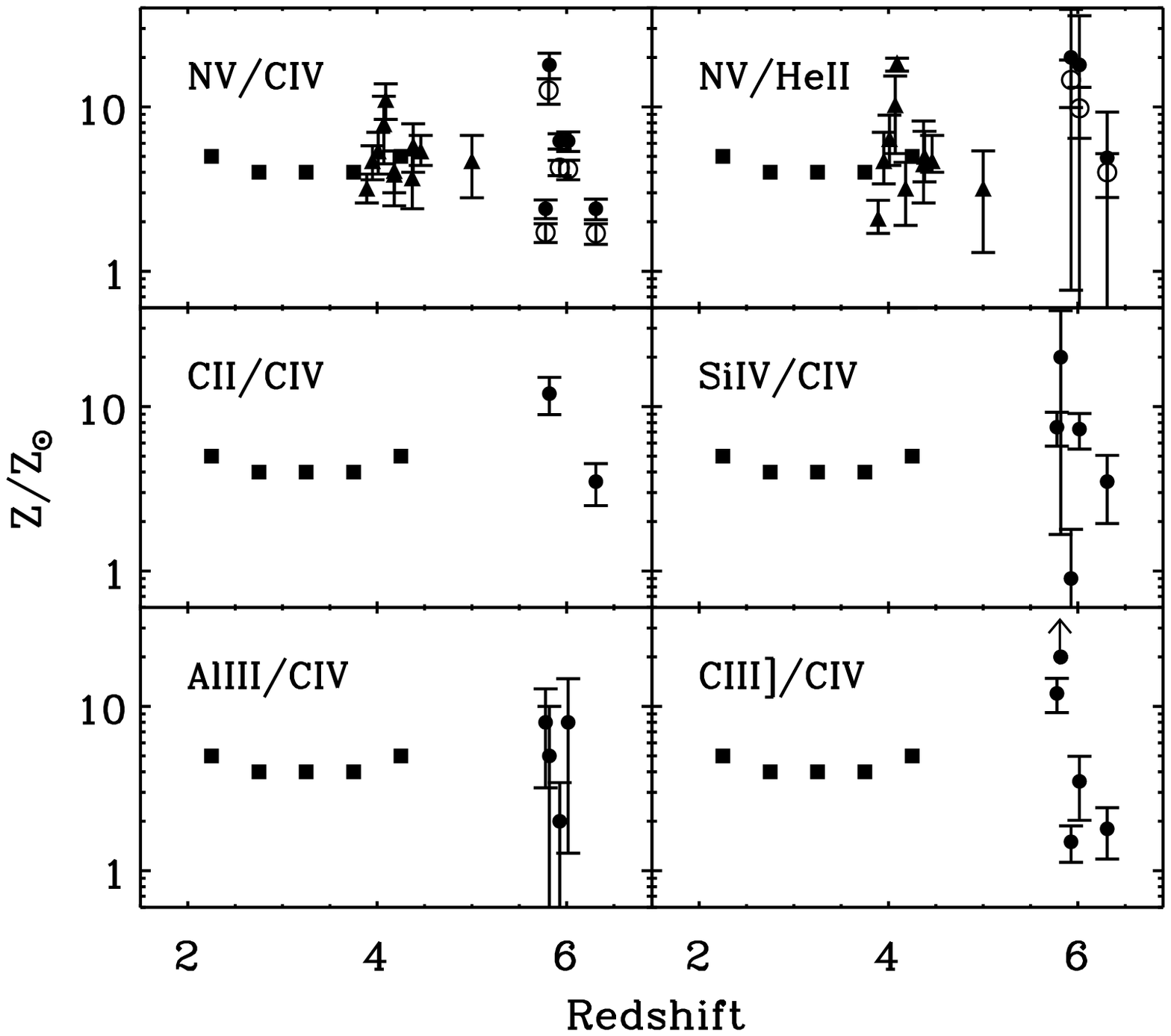}
\caption{Chemical abundances of quasars derived from line 
ratios as a function of redshift. The open circles represent the metallicities
derived from the \nv/\civ\ and \nv/\heii\ ratios using Figure 5 of 
\citet{ham02}. The filled circles represent the metallicities derived from the 
line ratios using Figure 29 of \citet{nag06}. The filled squares show the 
averaged metallicities of quasars for the luminosity range $-28.5<M_B<-25.5$ 
from \citet{nag06}. The filled triangles show the metallicities of eleven 
$3.9<z<5.0$ quasars from \citet{die03a}. The metallicities estimated from each 
line ratio show no strong evolution up to $z>6$ within the errors.}
\end{figure}

\clearpage
\begin{figure}
\epsscale{0.6}
\plotone{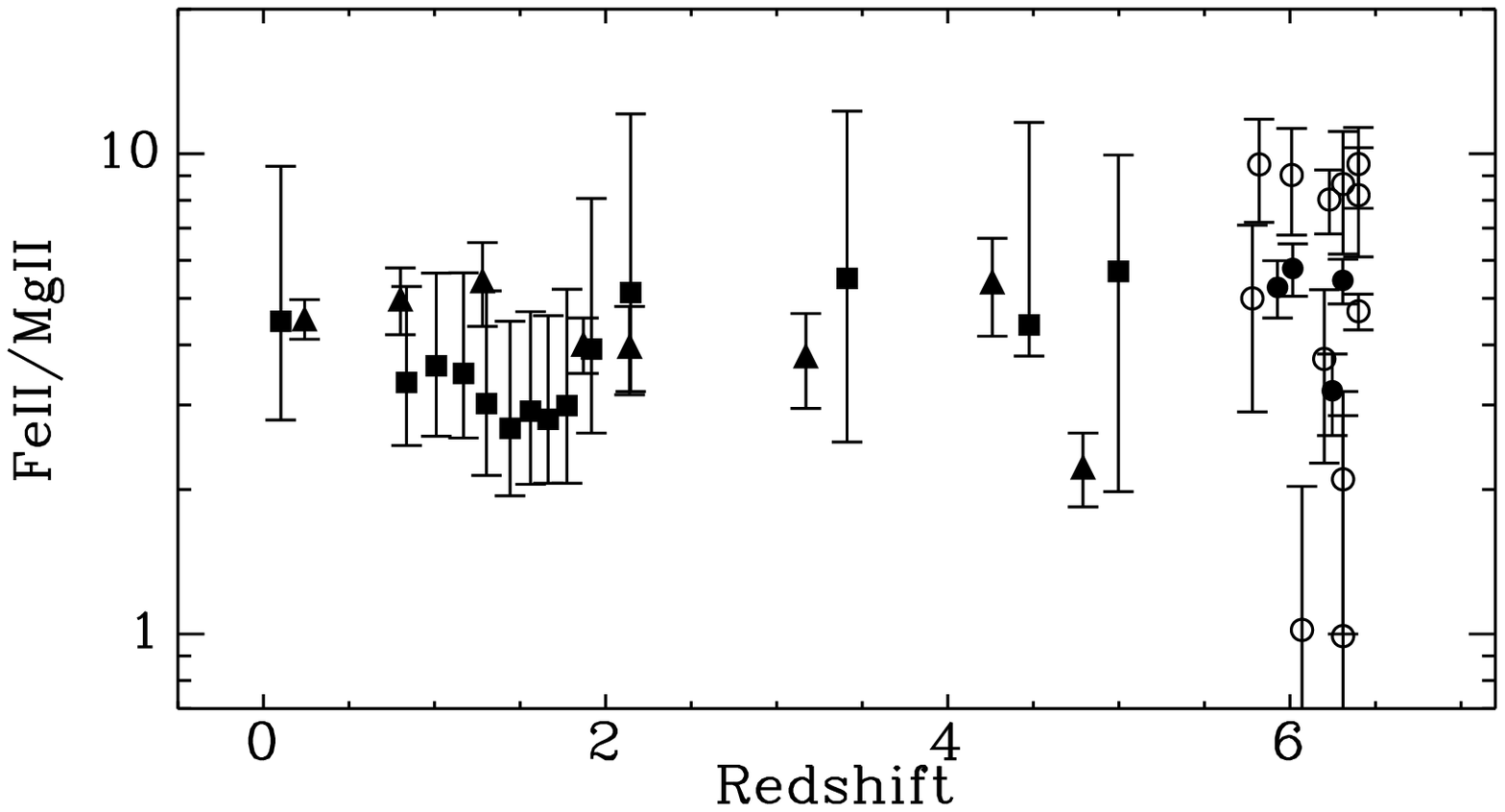}
\caption{The \feii/\mgii\ ratio as a function of redshift. The filled 
circles are our results. The open circles show the \feii/\mgii\ ratios of
$z\sim6$ quasars from previous measurements \citep{bar03,mai03,fre03,iwa04}. 
The filled squares and triangles represent the low-redshift results from 
\citet{iwa02} and \citet{die03b}. The lack of strong evolution of the 
\feii/\mgii\ abundance continues to $z\sim6$.}
\end{figure}

\clearpage
\begin{figure}
\epsscale{0.8}
\plotone{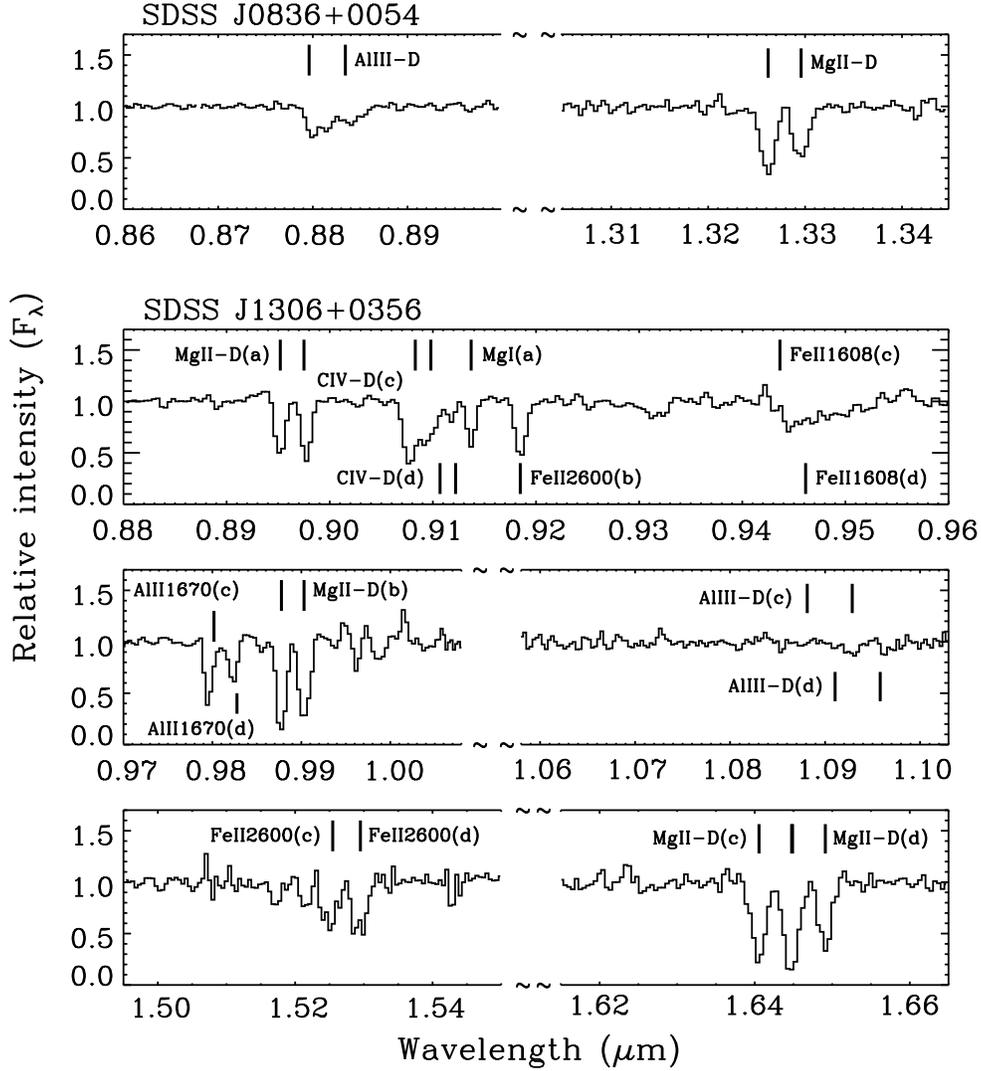}
\caption{Normalized spectra of the five absorption systems in the spectra
of SDSS J0836+0054 and SDSS J1306+0356. The positions of the identified lines
are marked as vertical lines, and the names of the lines are given nearby.
Doublets are expressed as `--D'. The four absorption systems in the spectra
SDSS J1306+0356 are denoted as (a), (b), (c), and (d).}
\end{figure}

\clearpage 
\begin{figure}
\epsscale{0.6}
\plotone{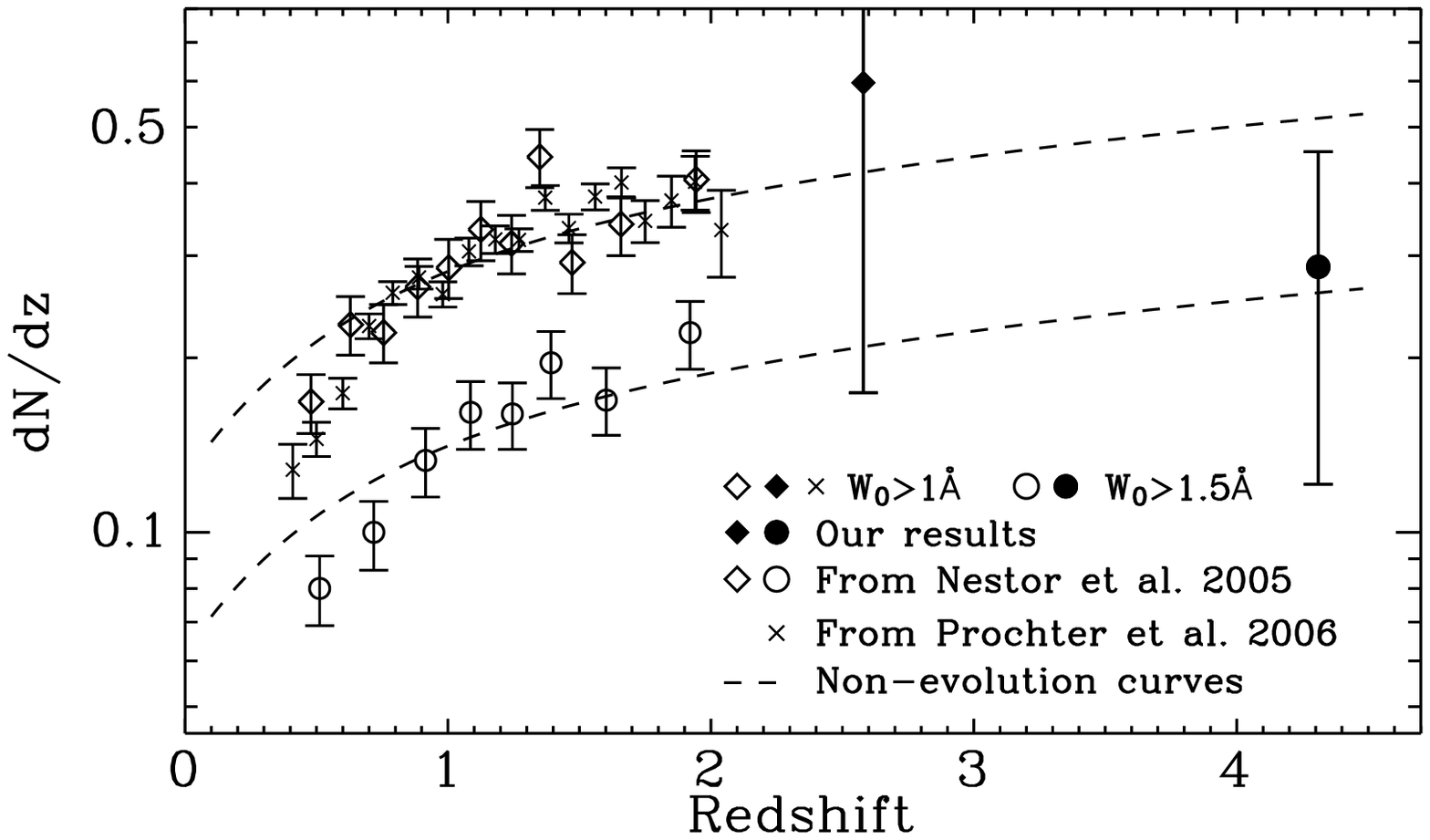} 
\caption{Comoving number densities of \mgii\ absorbers. The filled
diamond represents the density for $W_0^{\lambda2796}>1$ \AA\ at $z=2.6$ and
the filled circle represents the density for $W_0^{\lambda2796}>1.5$ \AA\
at $z=4.3$. The open diamonds and circles show the number densities for 
$W_0^{\lambda2796}>1.0$ and 1.5 \AA\ from Figure 9 of \citet{nes05}. The 
crosses show the densities of the $W_0^{\lambda2796}>1.0$ \AA\ absorbers from 
\citet{pro06}. The dashed lines represent non-evolution 
curves for the cosmology of $(\Omega_m, \Omega_{\Lambda}, h)=(0.3, 0.7, 0.7)$. 
The curves have been scaled to minimize the $\chi^2$ to the data of 
\citet{nes05}. The densities are consistent with no cosmic evolution.}
\end{figure} 

\clearpage
\begin{deluxetable}{ccccllc}
\rotate
\tablecaption{Log of Observations}
\tablewidth{0pt}
\tablehead{\colhead{Quasar (SDSS)} & \colhead{Redshift} & 
  \colhead{$z_{AB}$ (mag)} & \colhead{$J$ (mag) } & \colhead{Date} & 
  \colhead{Instrument} & \colhead{$t_{exp}$ (min)} }
\startdata
J083643.85+005453.3    & 5.82 & 18.74$\pm$0.05 & 17.89$\pm$0.05 
  & 2006 Feb 19     & GNIRS &  70  \\
J103027.10+052455.0    & 6.28 & 20.05$\pm$0.10 & 18.87$\pm$0.10 
  & 2006 Mar 20, 26 & GNIRS & 140  \\
J104433.04$-$012502.2  & 5.80 & 19.23$\pm$0.07 & 18.31$\pm$0.10
  & 2006 Mar 28     & GNIRS & 120  \\
J130608.26+035626.3    & 5.99 & 19.47$\pm$0.05 & 18.77$\pm$0.10 
  & 2006 Mar 26, 27 & GNIRS & 140  \\
J141111.29+121737.4    & 5.93 & 19.65$\pm$0.08 & 18.95$\pm$0.05
  & 2006 Mar 28     & GNIRS & 120  \\
J162331.81+311200.5    & 6.22 & 20.09$\pm$0.10 & 19.15$\pm$0.10
  & 2004 Aug 16, Sep 14 & NIRI (K) & 180 \\
  &      &       &       & 2005 Aug 22, 24  &         &     \\
\enddata
\tablecomments{Redshifts, $z_{AB}$ (AB magnitudes) and $J$ (Vega-based 
  magnitudes) are from the quasar discovery papers \citet{fan00}, 
  \citet{fan01}, and \citet{fan04}.}
\end{deluxetable}

\clearpage
\begin{deluxetable}{lcrc}
\tablecaption{Redshifts and Continuum Slopes}
\tablewidth{0pt}
\tablehead{\colhead{Quasar (SDSS)} & \colhead{Redshift} 
  & \colhead{$\rm Slope_{uv}$($\alpha_{\nu}$)\tablenotemark{a}} 
  & \colhead{$\rm Slope_{opt}$($\alpha_{\nu}$)\tablenotemark{b}}}
\startdata
J0836+0054 & $5.810\pm0.003$\tablenotemark{c}& $-0.62_{-0.06}^{+0.06}$ & $-0.52$\\
J1030+0524 & $6.309\pm0.009$                 & $0.46_{-0.25}^{+0.18}$  & $-0.31$\\
J1044$-$0125&$5.778\pm0.005$\tablenotemark{d}& $-0.27_{-0.10}^{+0.09}$ & $-0.30$\\
J1306+0356 & $6.016\pm0.005$                 & $0.50_{-0.14}^{+0.12}$  & $-0.12$\\
J1411+1217 & $5.927\pm0.004$                 & $-0.21_{-0.19}^{+0.16}$ & $-0.51$\\
J1623+3112 & $6.247\pm0.005$                 & $\ldots$                & $-0.32$\\
\enddata
\tablenotetext{a}{UV continuum slopes measured from the Gemini spectra.}
\tablenotetext{b}{Optical continuum slopes taken from \citet{jia06}.}
\tablenotetext{c}{Taken from \citet{kur07}.}
\tablenotetext{d}{Measured from the \ciii\ $\lambda$1909 emission line.}
\end{deluxetable}

\clearpage
\begin{deluxetable}{llllllll}
\tabletypesize{\small} 
\rotate
\tablecaption{Emission Line Properties}
\tablewidth{0pt}
\tablehead{\colhead{} & \colhead{} & 
 \colhead{J0836+0054} & \colhead{J1030+0524} & \colhead{J1044$-$0125} & 
 \colhead{J1306+0356} & \colhead{J1411+1217} & \colhead{J1623+3112}}
\startdata
\nv    & EW   & 15.9$\pm$0.7  & 4.9$\pm$0.1   & 3.9$\pm$0.2
              & 9.3$\pm$0.3   & 28.9$\pm$0.4  & $\ldots$ \\
       & FWHM & 31.5$\pm$1.0  & 8.7$\pm$0.2   & 10.4$\pm$0.3
              & 15.0$\pm$0.2  & 22.3$\pm$0.3  & $\ldots$ \\
\siii  & EW   & 3.6$\pm$0.3   & 3.0$\pm$0.2   & 2.3$\pm$0.1
              & 1.7$\pm$0.1   & 0.9$\pm$0.1   & $\ldots$ \\
       & FWHM & 15.0$\pm$0.6  & 15.8$\pm$0.7  & 12.7$\pm$0.3
              & 12.8$\pm$0.4  & 3.7$\pm$0.3   & $\ldots$ \\
\oi    & EW   & 7.4$\pm$0.1   & 6.7$\pm$0.3   & 2.6$\pm$0.2
              & $\ldots$      & 6.7$\pm$0.2   & $\ldots$ \\
       & FWHM & 24.6$\pm$0.5  & 18.8$\pm$0.7  & 24.3$\pm$1.3
              & $\ldots$      & 8.1$\pm$0.2   & $\ldots$ \\
\cii   & EW   & 5.8$\pm$0.5   & 4.0$\pm$0.4   & $\ldots$
              & $\ldots$      & $\ldots$      & $\ldots$ \\
       & FWHM & 41.9$\pm$3.1  & 17.0$\pm$1.5  & $\ldots$
              & $\ldots$      & $\ldots$      & $\ldots$ \\
\siiv  & EW   & 9.3$\pm$4.2   & 8.5$\pm$1.7   & 8.9$\pm$0.7
              & 9.5$\pm$0.8   & 8.1$\pm$6.4   & $\ldots$ \\
       & FWHM & 25.3$\pm$5.9  & 18.4$\pm$2.9  & 35.7$\pm$1.8
              & 29.0$\pm$2.0  & 21.5$\pm$12.2 & $\ldots$ \\
\civ   & EW   & 15.0$\pm$1.6  & 40.0$\pm$3.1  & 24.7$\pm$2.1
              & 29.6$\pm$2.4  & 76.2$\pm$5.8  & $\ldots$ \\
       & FWHM & 37.1$\pm$3.0  & 27.3$\pm$2.9  & 42.2$\pm$3.1
              & 24.5$\pm$2.3  & 17.5$\pm$1.6  & $\ldots$ \\
\heii  & EW   & $\ldots$      & 3.4$\pm$2.0   & $\ldots$
              & 2.6$\pm$1.8   & 5.1$\pm$2.5   & $\ldots$ \\
       & FWHM & $\ldots$      & 21.7$\pm$8.8  & $\ldots$
              & 15.5$\pm$7.5  & 39.0$\pm$26.8 & $\ldots$ \\
\oiii  & EW   & $\ldots$      & 2.3$\pm$1.3   & $\ldots$
              & 2.7$\pm$2.0   & 2.8$\pm$2.5   & $\ldots$ \\
       & FWHM & $\ldots$      & 22.9$\pm$11.0 & $\ldots$
              & 26.9$\pm$19.4 & 42.2$\pm$34.0 & $\ldots$ \\
\aliii & EW   & 1.8$\pm$0.9   & $\ldots$      & 5.0$\pm$1.6
              & 7.0$\pm$3.1   & 5.1$\pm$2.3   & $\ldots$ \\
       & FWHM & 16.3$\pm$6.8  & $\ldots$      & 31.0$\pm$6.7
              & 34.5$\pm$14.7 & 25.6$\pm$9.2  & $\ldots$ \\
\ciii  & EW   & 18.9$\pm$2.3  & 19.3$\pm$2.8  & 24.0$\pm$1.8
              & 19.1$\pm$3.9  & 26.1$\pm$3.2  & $\ldots$ \\
       & FWHM & 53.2$\pm$4.9  & 31.6$\pm$3.3  & 50.3$\pm$3.2
              & 37.6$\pm$7.4  & 31.3$\pm$2.6  & $\ldots$ \\
\mgii  & EW   & $\ldots$      & 46.1$\pm$3.7  & $\ldots$
              & 24.9$\pm$2.1  & 37.5$\pm$5.5  & 27.4$\pm$3.9 \\
       & FWHM & $\ldots$      & 33.8$\pm$2.2  & $\ldots$
              & 33.2$\pm$2.2  & 23.7$\pm$2.7  & 34.4$\pm$2.4 \\
\enddata
\tablecomments{Rest-frame FWHM and EW are in units of \AA.}
\end{deluxetable}

\clearpage
\begin{deluxetable}{lcccccc}
\tabletypesize{\small}
\rotate
\tablecaption{Flux Ratios}
\tablewidth{0pt}
\tablehead{\colhead{} &
 \colhead{J0836+0054} & \colhead{J1030+0524} & \colhead{J1044$-$0125} &
 \colhead{J1306+0356} & \colhead{J1411+1217} & \colhead{J1623+3112}}
\startdata
\nv/\civ    &  1.44$\pm$0.17  &  0.21$\pm$0.02  &  0.23$\pm$0.03  
            &  0.55$\pm$0.05  &  0.56$\pm$0.04  &  $\ldots$ \\
\siii/\civ  &  0.32$\pm$0.04  &  0.12$\pm$0.02  &  0.13$\pm$0.02 
            &  0.10$\pm$0.01  &  0.02$\pm$0.01  &  $\ldots$ \\
\oi/\civ    &  0.62$\pm$0.07  &  0.26$\pm$0.02  &  0.14$\pm$0.02 
            &  $\ldots$       &  0.12$\pm$0.02  &  $\ldots$ \\
\cii/\civ   &  0.47$\pm$0.06  &  0.14$\pm$0.02  &  $\ldots$
            &  $\ldots$       &  $\ldots$       &  $\ldots$ \\
\siiv/\civ  &  0.72$\pm$0.33  &  0.27$\pm$0.06  &  0.43$\pm$0.05
            &  0.41$\pm$0.05  &  0.13$\pm$0.10  &  $\ldots$ \\
\heii/\civ  &  $\ldots$       &  0.07$\pm$0.04  &  $\ldots$
            &  0.08$\pm$0.05  &  0.06$\pm$0.05  &  $\ldots$ \\
\oiii/\civ  &  $\ldots$       &  0.05$\pm$0.03  &  $\ldots$
            &  0.08$\pm$0.06  &  0.03$\pm$0.02  &  $\ldots$ \\
\aliii/\civ &  0.10$\pm$0.05  &  $\ldots$       &  0.15$\pm$0.05
            &  0.15$\pm$0.07  &  0.05$\pm$0.02  &  $\ldots$ \\
\ciii/\civ  &  0.95$\pm$0.15  &  0.29$\pm$0.05  &  0.68$\pm$0.08
            &  0.38$\pm$0.08  &  0.24$\pm$0.03  &  $\ldots$ \\
\feii/\mgii &  $\ldots$       &  5.45$\pm$0.58  &  $\ldots$ 
            &  5.77$\pm$0.72  &  5.27$\pm$0.72  &  3.21$\pm$0.62 \\
\enddata
\end{deluxetable}

\clearpage
\begin{deluxetable}{cccccc}
\tablecaption{Central BH Masses ($10^9$ M$_{\sun}$)}
\tablewidth{0pt}
\tablehead{\colhead{Quasar (SDSS)} &\colhead{$\rm L_{Bol}$\tablenotemark{a}} &
  \colhead{$\rm L_{Bol}/L_{Edd}$\tablenotemark{b}} & 
  \colhead{$\rm M_{BH}$(\civ)} & \colhead{$\rm M_{BH}$(\mgii)} & 
  \colhead{$\rm M_{BH}^{'}$(\mgii)\tablenotemark{c}} }
\startdata
J0836+0054   & 47.72 & 0.44 & 9.3$\pm$1.6  & $\ldots$    & $\ldots$    \\
J1030+0524   & 47.37 & 0.50 & 3.6$\pm$0.9  & 1.0$\pm$0.2 & 2.1$\pm$0.4 \\
J1044$-$0125 & 47.63 & 0.31 & 10.5$\pm$1.6 & $\ldots$    & $\ldots$    \\
J1306+0356   & 47.40 & 0.61 & 3.2$\pm$0.6  & 1.1$\pm$0.1 & 2.2$\pm$0.3 \\
J1411+1217   & 47.20 & 0.94 & 1.3$\pm$0.3  & 0.6$\pm$0.1 & 0.9$\pm$0.2 \\
J1623+3112   & 47.33 & 1.11 & $\ldots$     & 1.5$\pm$0.3 & $\ldots$    \\
\enddata
\tablenotetext{a}{Bolometric luminosity in log[erg s$^{-1}$] from 
  \citet{jia06}.}
\tablenotetext{b}{$\rm L_{Edd}$ is derived from $\rm M_{BH}$(\civ) except for 
  SDSS J1623+3112, whose $\rm L_{Edd}$ is derived from $\rm M_{BH}$(\mgii).}
\tablenotetext{c}{$\rm M_{BH}^{'}$ is estimated from the new relation by 
  Vestergaard et al. (in preparation).}
\end{deluxetable}

\clearpage
\begin{deluxetable}{lrrcc}
\tablecaption{Absorption Lines}
\tablewidth{0pt}
\tablehead{\colhead{Absorption system} & \colhead{$\lambda_{abs}$ (\AA)} 
  & \colhead{ID} & \colhead{$z_{abs}$} & \colhead{$W_0$ (\AA)}}
\startdata
SDSS J0836+0054    & 13261.81 &  \mgii(2796) &  3.7426 &   2.59\\
                   & 13295.43 &  \mgii(2803) &  3.7424 &   2.08\\
                   & & & & \\
SDSS J1306+0356 (a)&  8952.03 &  \mgii(2796) &  2.2013 &   1.86\\
                   &  8977.30 &  \mgii(2803) &  2.2022 &   1.99\\
                   &  9137.13 &  \mgit(2852) &  2.2028 &   1.29\\
                   & & & & \\
SDSS J1306+0356 (b)&  9877.64 &  \mgii(2796) &  2.5323 &   3.24\\
                   &  9903.27 &  \mgii(2803) &  2.5324 &   3.15\\
                   &  9185.23 &  \feii(2600) &  2.5325 &   1.94\\
                   & & & & \\
SDSS J1306+0356 (c)& 16405.55 &  \mgii(2796) &  4.8668 &   2.92\\
                   & 16447.67 &  \mgii(2803) & $\ldots$&  $\ldots$\\
                   &  9795.84 &  \aliit(1670)&  4.8632 &   1.11\\
                   & 15252.21 &  \feii(2600) &  4.8658 &   1.49\\
                   & & & & \\
SDSS J1306+0356 (d)& 16448.85 &  \mgii(2796) & $\ldots$&  $\ldots$\\
                   & 16491.08 &  \mgii(2803) &  4.8823 &   2.11\\
                   &  9822.93 &  \aliit(1670)&  4.8795 &   0.76\\
                   & 15293.94 &  \feii(2600) &  4.8819 &   1.97\\
\enddata
\end{deluxetable}


\begin{thebibliography}{}
\bibitem[Abazajian et al.(2004)]{aba04} Abazajian, K., et al.\ 2004, 
  \aj, 128, 502 
\bibitem[Baldwin et al.(1995)]{bal95} Baldwin, J., Ferland, G., Korista, K., 
  \& Verner, D.\ 1995, \apjl, 455, L119
\bibitem[Baldwin et al.(2003)]{bal03} Baldwin, J.~A., Hamann, F., Korista, 
  K.~T., Ferland, G.~J., Dietrich, M., \& Warner, C.\ 2003, \apj, 583, 649
\bibitem[Baldwin et al.(2004)]{bal04} Baldwin, J.~A., Ferland, G.~J., 
  Korista, K.~T., Hamann, F., \& LaCluyz{\'e}, A.\ 2004, \apj, 615, 610 
\bibitem[Barth et al.(2003)]{bar03} Barth, A.~J., Martini, P., Nelson, C.~H.,
  \& Ho, L.~C. 2003, \apj, 594, L95
\bibitem[Becker et al.(1995)]{bec95} Becker, R.~H., White, R.~L.,
    \& Helfand, D.~J. 1995, \apj, 450, 559
\bibitem[Bechtold et al.(2003)]{bec03} Bechtold, J., et al. 2003, \apj,
  588, 119
\bibitem[Bertoldi et al.(2003a)]{ber03a} Bertoldi, F., et al. 2003, \aap, 406,
  L55
\bibitem[Bertoldi et al.(2003b)]{ber03b} Bertoldi, F., et al. 2003, \aap, 409,
  L47
\bibitem[Carilli et al.(2001)]{car01} Carilli, C. L., Bertoldi, F., Omont, A.,
  Cox, P., McMahon, R. G., \& Isaak, K. G. 2001, \aj, 122, 1679
\bibitem[Charlton \& Churchill(1998)]{cha98} Charlton, J.~C., \& 
  Churchill, C.~W.\ 1998, \apj, 499, 181
\bibitem[Churchill et al.(2005)]{chu05} Churchill, C. W., Kacprzak, G. G.,
  \& Steidel, C. C. 2005, Proceedings IAU Colloquium No. 199, Probing
  Galaxies through Quasar Absorption Lines, P. R. Williams, C. Shu, and
  B. M\'enard, eds., astro-ph/0504392
\bibitem[Churchill et al.(2000)]{chu00} Churchill, C.~W., Mellon, R.~R., 
  Charlton, J.~C., Jannuzi, B.~T., Kirhakos, S., Steidel, C.~C., \& 
  Schneider, D.~P.\ 2000, \apj, 543, 577
\bibitem[Cool et al.(2006)]{coo06} Cool, R.~J., et al.\ 2006, \aj, 132, 823
\bibitem[Croton et al.(2006)]{cro06} Croton, D. J., et al. 2006, \mnras,
    365, 11
\bibitem[Dietrich et al.(2002)]{die02} Dietrich, M., Appenzeller, I., 
  Vestergaard, M., \& Wagner, S. J. 2002, \apj, 564, 581
\bibitem[Dietrich et al.(2003a)]{die03a} Dietrich, M., Appenzeller, I., 
  Hamann, F., Heidt, J., J{\"a}ger, K., Vestergaard, M., 
  \& Wagner, S.~J.\ 2003, \aap, 398, 891
\bibitem[Dietrich et al.(2003b)]{die03b} Dietrich, M., Hamann, F., Appenzeller,
  I., \& Vestergaard, M. 2003, \apj, 596, 817
\bibitem[Dietrich \& Hamann(2004)]{die04} Dietrich, M., \& Hamann, F. 2004,
  \apj, 611, 761
\bibitem[Djorgovski et al.(2001)]{djo01} Djorgovski, S. G., Castro, S.,
  Stern, D., \& Mahabal, A. A. 2001, \apj, 560, L5
\bibitem[Elston et al.(1996)]{els96} Elston, R., Bechtold, 
  J., Hill, G.~J., \& Ge, J.\ 1996, \apj, 456, L13 
\bibitem[Elvis et al.(1994)]{elv94} Elvis, M., et al. 1994, \apjs, 95, 1
\bibitem[Fan et al.(2000)]{fan00} Fan, X., et al. 2000, \aj, 120, 1167
\bibitem[Fan et al.(2001)]{fan01} Fan, X., et al. 2001, \aj, 122, 2833
\bibitem[Fan et al.(2003)]{fan03} Fan, X., et al. 2003, \aj, 125, 1649
\bibitem[Fan et al.(2004)]{fan04} Fan, X., et al. 2004, \aj, 128, 515
\bibitem[Fan et al.(2006)]{fan06} Fan, X., et al. 2006, \aj, 131, 1203
\bibitem[Fechner et al.(2004)]{fec04} Fechner, C., Baade, R., 
  \& Reimers, D.\ 2004, \aap, 418, 857 
\bibitem[Foltz et al.(1987)]{fol87} Foltz, C.~B., Weymann, R.~J., Morris, 
  S.~L., \& Turnshek, D.~A.\ 1987, \apj, 317, 450 
\bibitem[Freudling et al.(2003)]{fre03} Freudling, W., Corbin,
  M. R., \& Korista, K. T. 2003, \apj, 587, L67
\bibitem[Friaca \& Terlevich(1998)]{fri98} Friaca, A.~C.~S., 
  \& Terlevich, R.~J.\ 1998, \mnras, 298, 399 
\bibitem[Goodrich et al.(2001)]{goo01} Goodrich, R. W., et al. 2001, \apj,
  561, L23
\bibitem[Greggio \& Renzini(1983)]{gre83} Greggio, L., \& Renzini, A. 1983,
  \aap, 118, 217
\bibitem[Hamann \& Ferland(1992)]{ham92} Hamann, F., \& Ferland, G. 1992,
  \apj, 391, L53
\bibitem[Hamann \& Ferland(1993)]{ham93} Hamann, F., \& Ferland, G. 1993,
  \apj, 418, 11
\bibitem[Hamann \& Ferland(1999)]{ham99} Hamann, F., \& Ferland, G. 1999,
  \araa, 37, 487
\bibitem[Hamann et al.(2002)]{ham02} Hamann, F., Korista, K.~T., Ferland, 
  G.~J., Warner, C., \& Baldwin, J.\ 2002, \apj, 564, 592 
\bibitem[Hamann et al.(2007)]{ham07} Hamann, F., Warner, C., Dietrich, M.,
  \& Ferland, G. 2007, in ASP Conference Proceedings, the Central Engine of 
  Active Galactic Nuclei, ed. L. C. Ho and J.-M. Wang (San Francisco: ASP), 
  in press (astro-ph/0701503)
\bibitem[Heger \& Woosley(2002)]{heg02} Heger, A., \& 
  Woosley, S.~E.\ 2002, \apj, 567, 532 
\bibitem[Hopkins et al.(2005)]{hop05} Hopkins, P. F., Hernquist, L., 
  Cox, T. J., Di Matteo, T., Martini, P., Robertson, B., \& Springel, V.
  2005, \apj, 630, 705
\bibitem[Hopkins et al.(2006)]{hop06} Hopkins, P.~F., Hernquist, L., Cox, 
  T.~J., Di Matteo, T., Robertson, B., \& Springel, V.\ 2006, \apjs, 163, 1 
\bibitem[Iwamuro et al.(2002)]{iwa02} Iwamuro, F., Motohara, K., Maihara, 
  T., Kimura, M., Yoshii, Y., \& Doi, M.\ 2002, \apj, 565, 63 
\bibitem[Iwamuro et al.(2004)]{iwa04} Iwamuro, F., Kimura, M., Eto, S.,
  Maihara, T., Motohara, K., Yoshii, Y., \& Doi, M. 2004, \apj, 614, 69
\bibitem[Jiang et al.(2006)]{jia06} Jiang, L., et al. 2006, \aj, 132, 2127
\bibitem[Kauffmann \& Haehnelt(2000)]{kau00} Kauffmann, G., \& Haehnelt, M.
    2000, \mnras, 311, 576
\bibitem[Kollmeier et al.(2006)]{kol06} Kollmeier, J.~A., et al.\ 2006,
  \apj, 648, 128
\bibitem[Kurk et al.(2007)]{kur07} Kurk, J.~D. et al.\ 2007, \apj, submitted
\bibitem[Li et al.(2007)]{li07} Li, Y., et al. 2007, \apj, in press
  (astro-ph/0608190)
\bibitem[Lodato \& Natarajan(2007)]{lod07} Lodato, G., \&   Natarajan, P.\ 
  2007, \mnras, 377, L64
\bibitem[Madau \& Rees(2001)]{mad01} Madau, P., \& Rees, M.~J.\ 2001, \apjl, 
  551, L27
\bibitem[Maiolino et al.(2003)]{mai03} Maiolino, R., Juarez, Y., Mujica, R.,
  Nagar, N. M., \& Oliva, E. 2003, \apj, 596, L155
\bibitem[Matteucci \& Recchi(2001)]{mat01} Matteucci, F., \& 
  Recchi, S.\ 2001, \apj, 558, 351 
\bibitem[McGreer et al.(2006)]{mcg06} McGreer, I.~D., Becker, R.~H.,
  Helfand, D.~J., \& White, R.~L.\ 2006, \apj, 652, 157
\bibitem[McIntosh et al.(1999)]{mci99} McIntosh, D. H., Rix, H. W., Rieke,
  M. J., \& Foltz, C. B. 1999, \apj, 517, L73
\bibitem[McLure \& Dunlop(2004)]{mcl04} McLure, R.~J., \&
  Dunlop, J.~S.\ 2004, \mnras, 352, 1390
\bibitem[Nagao et al.(2006)]{nag06} Nagao, T., Marconi, A., 
  \& Maiolino, R.\ 2006, \aap, 447, 157
\bibitem[Nestor et al.(2005)]{nes05} Nestor, D.~B., Turnshek, 
  D.~A., \& Rao, S.~M.\ 2005, \apj, 628, 637 
\bibitem[Pentericci et al.(2002)]{pen02} Pentericci, L., et al. 2002, \aj,
  123, 2151
\bibitem[Pentericci et al.(2003)]{pen03} Pentericci, L., et al. 2003, \aap,
  410, 75
\bibitem[Petric et al.(2003)]{pet03} Petric, A. O., et al. 2003, \aj, 126, 15
\bibitem[Priddey et al.(2003)]{pri03} Priddey, R. S., Isaak, K. G., McMahon,
  R. G., Robson, E. I., \& Pearson, C. P. 2003, \mnras, 344, L74
\bibitem[Prochter et al.(2006)]{pro06} Prochter, G.~E., Prochaska, J.~X., 
  \& Burles, S.~M.\ 2006, \apj, 639, 766 
\bibitem[Rees \& Volonteri(2006)]{ree06} Rees, M.~J., \& Volonteri, M.\ 2006, 
  Proceedings IAU Symposium No. 238, Black Holes: from Stars to Galaxies --
  across the Range of Masses, V. Karas and G. Matt, eds., astro-ph/0701512
\bibitem[Richards et al.(2002)]{ric02} Richards, G. T., et al. 2002,
  \aj, 124, 1
\bibitem[Robson et al.(2004)]{rob04} Robson, I., Priddey, R. S., Isaak, K. G.,
  \& McMahon, R. G. 2004, \mnras, 351, L29
\bibitem[Shemmer et al.(2006)]{she06} Shemmer, O., et al. 2006, \apj,
  644, 86
\bibitem[Shen et al.(2007)]{she07} Shen, Y., et al.\ 2007, \aj, 133, 2222 
\bibitem[Spergel et al.(2007)]{spe07} Spergel, D.~N., et al.\ 2007, \apjs, 
  170, 377 
\bibitem[Steidel et al.(2002)]{ste02} Steidel, C.~C., Kollmeier, J.~A.,
  Shapley, A.~E., Churchill, C.~W., Dickinson, M., \& Pettini, M.\ 2002,
  \apj, 570, 526
\bibitem[Steidel \& Sargent(1992)]{ste92} Steidel, C.~C., \& 
  Sargent, W.~L.~W.\ 1992, \apjs, 80, 1
\bibitem[Stern et al.(2003)]{ste03} Stern, D., et al. 2003, \apj,
  596, L39
\bibitem[Stoughton et al.(2002)]{sto02} Stoughton, C., et al.\ 2002, 
  \aj, 123, 485 
\bibitem[Trump et a.(2006)]{tru06} Trump, J.~R., et al. 2006, \apjs, 165, 1
\bibitem[Vanden Berk et al.(2001)]{van01} Vanden Berk, D. E., et al. 2001,
    \aj, 122, 549
\bibitem[Venkatesan et al.(2004)]{ven04} Venkatesan, A., Schneider, R., \& 
  Ferrara, A.\ 2004, \mnras, 349, L43 
\bibitem[Vestergaard \& Wilkes(2001)]{ves01} Vestergaard, M. \& Wilkes, B. J.
  2001, \apjs, 134, 1
\bibitem[Vestergaard(2004)]{ves04} Vestergaard, M. 2004, \apj, 601, 676
\bibitem[Vestergaard \& Peterson(2006)]{ves06} Vestergaard, M., \&
  Peterson, B.~M.\ 2006, \apj, 641, 689
\bibitem[Vilkoviskij \& Irwin(2001)]{vil01} Vilkoviskij, E.~Y., \& 
  Irwin, M.~J.\ 2001, \mnras, 321, 4 
\bibitem[Volonteri \& Rees(2006)]{vol06} Volonteri, M., \& Rees, M.~J.\ 2006, 
  \apj, 650, 669 
\bibitem[Walter et al.(2003)]{wal03} Walter, F., et al. 2003, Nature, 424, 406
\bibitem[Walter et al.(2004)]{wal04} Walter, F., et al. 2004, \apj, 615, L17
\bibitem[Wandel et al.(1999)]{wan99} Wandel, A., Peterson, B.~M., \& 
  Malkan, M.~A.\ 1999, \apj, 526, 579 
\bibitem[Wang et al.(2007)]{wan07} Wang, R., et al. \aj, in press
  (astro-ph/0704.2053)
\bibitem[Wyithe \& Loeb(2003)]{wyi03} Wyithe, J. S. B., \& Loeb, A. 2003,
    \apj, 595, 614
\bibitem[York et al.(2000)]{yor00} York, D. G., et al. 2000, \aj, 120, 1579
\end{thebibliography}
\end{document}